\documentclass[review]{elsarticle}

\usepackage{amssymb}
\usepackage[figuresright]{rotating}
\usepackage{multirow}
\usepackage{algorithm}
\usepackage{algorithmicx}
\usepackage{algpseudocode}
\usepackage{subfigure}
\usepackage{amsmath}
\usepackage{amsthm}
\usepackage{amsfonts}
\usepackage{bbm}
\usepackage{color}
\usepackage{url}
\usepackage{booktabs}
\usepackage{ulem}

\newtheorem{theorem}{Theorem}
\newtheorem{lemma}{Lemma}
\newtheorem{definition}{Definition}
\newtheorem{example}{Example}

\usepackage{lineno}

\journal{Information Sciences}

\begin{document}

\begin{frontmatter}



\title{Sensitivity estimation for differentially private query processing}


\author[label1]{Meifan Zhang}
\author[label1]{Xin Liu}
\author[label1]{Lihua Yin$^*$}
\cortext[mycorrespondingauthor]{Corresponding author}
\ead[url]{yinlh@gzhu.edu.cn}

\affiliation[label1]{organization={Cyberspace Institute of Advanced Technology},
            addressline={Guangzhou University},
            city={Guangzhou},
            postcode={510006},
            state={},
            country={China}}

%

\begin{abstract}

Differential privacy has become a popular privacy-preserving method in data analysis, query processing, and machine learning, which adds noise to the query result to avoid leaking privacy. Sensitivity, or the maximum impact of deleting or inserting a tuple on query results, determines the amount of noise added. Computing the sensitivity of some simple queries such as counting query is easy, however, computing the sensitivity of complex queries containing join operations is challenging. Global sensitivity of such a query is unboundedly large, which corrupts the accuracy of the query answer. Elastic sensitivity and residual sensitivity offer upper bounds of local sensitivity to reduce the noise, but they suffer from either low accuracy or high computational overhead. We propose two fast query sensitivity estimation methods based on sampling and sketch respectively, offering competitive accuracy and higher efficiency compared to the state-of-the-art methods.

\end{abstract}



\end{frontmatter}



\section{Introduction}\label{sec:introduction}
\begin{figure}[htbp]
\centering
\includegraphics[scale=0.4]{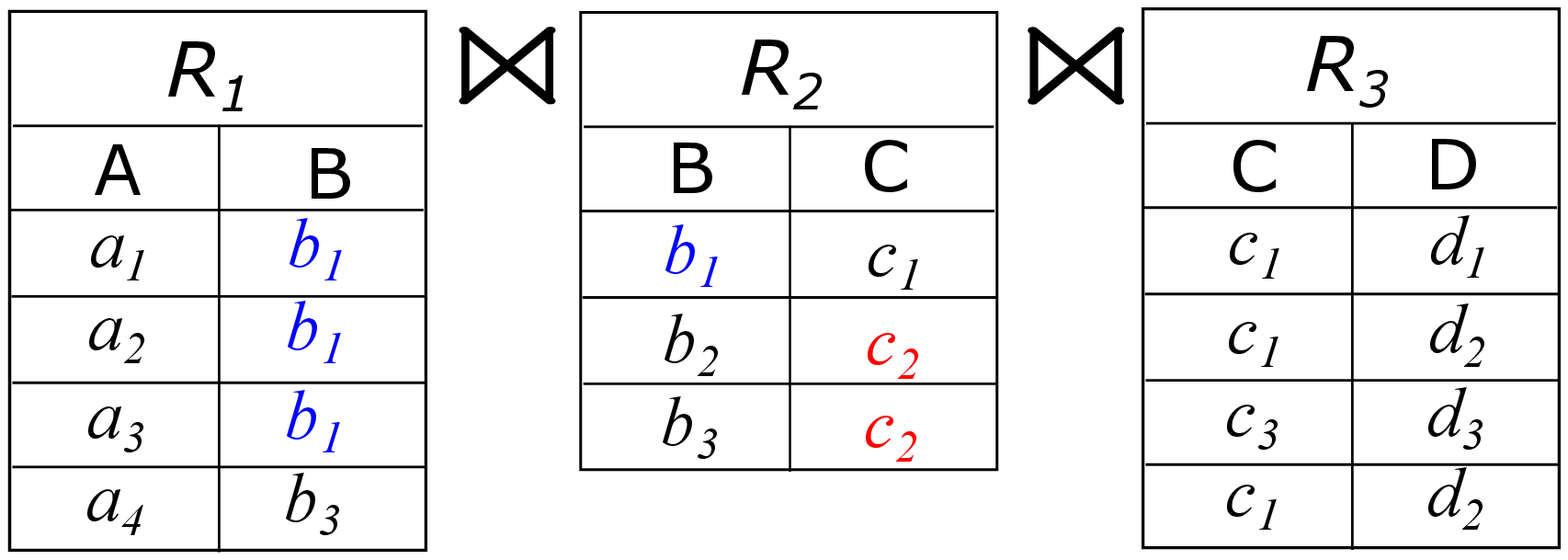}
\caption{Impact of deleting a tuple on the join size.}
\label{Fig:intro_example}
\end{figure}

Differential privacy provides strong privacy guarantees for query processing by adding random noise to the query answer. Usually, the noise is determined by the sensitivity of a query, which refers to the maximum difference between the query results on two datasets that differ by only one record. For example, the sensitivity of a query ``SELECT Count(*) FROM data WHERE Salary $>$ 5000;'' is 1, since inserting or removing one tuple has an effect on the result of at most 1. However, the sensitivity of a query containing join operators is more complex, since adding or removing a tuple may have significant influence upon the query result. Considering the query $q$=$Count(R_1(A,B)\Join R_2(B,C)\Join R_3(C,D))$ on the relations shown in Fig~\ref{Fig:intro_example}, it is difficult to calculate the sensitivity of such a query while ensuring the privacy, accuracy and efficiency. The global sensitivity of such a query is unbounded, because it depends on the query rather than the data. The local sensitivity depending on both the data and the query introduces less error but dose not satisfy the differential privacy. The state-of-the-art definitions of sensitivities such as the Elastic Sensitivity(ES)~\cite{Johnson2017TowardsPD} and the Residual Sensitivity(RS)~\cite{Dong2021ResidualSF} use the smooth upper bound of local sensitivity to solve this problem, but the accuracy and efficiency remain unsatisfactory. We use the following example to show the main ideas and shortcomings of ES and RS.

\begin{example}\label{example:intro}
   A join query $q$=$Count(R_1(A,B)\Join R_2(B,C)\Join R_3(C,D))$ is shown in Fig~\ref{Fig:intro_example}. ES and RS use different ways to compute the upper bound of the local sensitivity when deleting or inserting a tuple to the relations.

 (1) ES calculates the upper bound based on the maximum frequency of each join attribute as follows.
  \begin{align*}
    max( & mf(R_1.B)\cdot mf(R_2.C) \\
         &  mf(R_1.B)\cdot mf(R_3.C)  \\
         &  mf(R_2.B)\cdot mf(R_3.C) )
  \end{align*}
  The $mf(X)$ denotes the frequency of the most-frequent value of attribute $X$. The $mf(R_1.B)\cdot mf(R_2.C)$ means the upper bound for the frequency of the attribute $C$ in $R_1\Join R_2$. The worst-case sensitivity occurs when each tuple in $R_2$ with the most frequent value for $(R_2.C)$ also contains attribute value $(R_2.B)$ matching the most frequent value of  $(R_1.B)$. But the reality is not always the worst case, as shown in Fig~\ref{Fig:intro_example}, the most frequent value of  $(R_1.B)$ is $b_1$, which does not match the most frequent value $c_2$ of $(R_2.C)$. Therefore, the upper bound is much higher than the actual influence of inserting or deleting a tuple from $R_3$.

 (2) RS calculates the upper bound of the local sensitivity according to a list of group-by queries. For example, to compute the impact of adding one tuple from $R_3$ on the query result, RS calculates a statistic called  ``maximum boundary'' of a residual query $R_1\Join R_2$ of $q$, which means the maximum frequency of attribute $R_2.C$ in $R_1\Join R_2$. Since $R_2.C$ can join with $R_3$, the ``maximum boundary'' of $R_1\Join R_2$ determines the maximum influence of adding a tuple to $R_3$. Such a ``maximum boundary'' can be computed by the following query.

  $Q_1$: SELECT MAX(count) from (SELECT R2.C, COUNT(*) from R1,R2 where R1.B=R2.B GROUP BY R2.C) as T;

  It is much more accurate to estimate the influence of deleting or inserting a tuple to $R_3$, since it considers the true join result of the sub-query $R_1\Join R_2$. However, such a query is still expensive, since it contains ``Join'' and ``Group-By'' operators. The computation cost is even greater, as we need to calculate the impact of inserting or deleting a tuple from each table in the join query $q$. That is, computing RS requires many group-by queries in the form of $Q_1$.
\end{example}

\begin{figure}[htbp]
\centering
\includegraphics[scale=0.4]{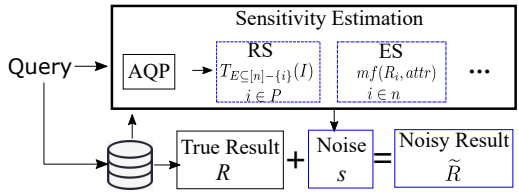}
\caption{Framework}
\label{Fig:framework}
\end{figure}

In general, ES can be efficiently computed according to the frequencies of join values, but has high errors. RS is more accurate but less efficient, since it is based on the true results of some sub-queries. As shown in~\cite{Dong2021ResidualSF}, computing the noisy multi-way join result according to RS takes nearly 10 times longer to query processing, which is too inefficient to support a fast response.
No existing work computes the sensitivity of a multi-way join query efficiently, while involving less error.
Our observation is that the query sensitivity is usually computed based on some statistics of the database, and more accurate sensitivity relies on more complex and time-consuming statistics.
To tackle this problem, we raise the idea of using approximate query processing (AQP) methods to approximately estimate the query sensitivity. We proposed a sensitivity estimation framework as shown in Fig~\ref{Fig:framework}. In this paper, we focus on sensitivity estimation for multi-way join queries. But the basic idea of approximately estimating the query sensitivity can be used for a variety of queries.

Estimating the query sensitivity remains non-trivial. On the one hand, simply using sampling methods to estimate the group-by queries involve unnecessary computation cost. Because a group-by query is usually conducted by estimating aggregations for each group, but as shown in figure~\ref{Fig:intro_example}, the aggregation function of such a group-by query for RS is MAX(COUNT), meaning that to estimate groups other than the largest group does not contribute to the result. Therefore, we need to focus on the estimation for the largest group. On the other hand, RS requires computing many sub-queries still contain join operators, which is still time-consuming even if sampling is adopted for each sub-query. Therefore, we proposed another algorithm to estimate all the required statistics while using less samples.

Even sampling methods can reduce the cost of sensitivity computation, drawing samples online is still costly. To tackle this problem, we proposed another sensitivity estimation method based on sketches. Sketches are useful for summarizing streams, they can be constructed with one-pass of data. We proposed sketch-based sensitivity estimation method, which constructs a sketch for each of the relations offline, and uses the sketches to estimate the sensitivity for online queries in $O(1)$.

This work makes the following contributions:
\begin{itemize}
  \item We propose a sampling-based sensitivity estimation method, which estimates the frequencies of the maximum groups in the residual queries of a multi-join query according to random walk. It improves the efficiency while remaining similar accuracy to the RS.
  \item We propose a sketch-based sensitivity estimation method. Sketching sensitivity is defined according to the AGMS sketch. Our sketch-based method is as efficient as ES, but much more accurate than ES. 
  \item Our experimental results show that both methods are more efficient than computing RS and ES while maintaining comparable accuracy.
\end{itemize}

The rest of the paper is organized as follows. We review the related work in Section 2. In Section 3 we introduce the preliminaries of differential privacy and existing definitions of sensitivity. Section 4 introduces two sensitivity estimation methods for multi-join queries based on sampling and sketches. In Section 5, we present an experimental comparison of our methods and existing ones. Section 6 concludes the paper.

\section{Related Works}

Differential Privacy~\cite{Dwork2006DifferentialP} is a mathematical framework to quantifying and managing the privacy risks. It is widely used in the field of privacy preserving data releasing~\cite{Aydre2021DifferentiallyPQ, Wang2020ContinuousRO} and mining~\cite{Maruseac2020PrecisionEnhancedDM, Wang2018LocallyDP}, machine learning~\cite{Triastcyn2019BayesianDP, Zheng2020ProtectingDB} and social network analysis~\cite{Jiang2023ApplicationsOD}.
Differential privacy can be easily applied to protect the privacy of query processing by adding noise correlated with the sensitivity to the query results~\cite{Dwork2006CalibratingNT} .

Computing the sensitivity of join queries is challenging.
Several mechanisms to support join queries have been proposed such as PINQ~\cite{McSherry2009PrivacyIQ} and wPINQ~\cite{Proserpio2012CalibratingDT}, but both of them are using global sensitivity, which means that the sensitivity could be extremely large thus resulting in low utility. Nissim et al.~\cite{Nissim2007SmoothSA} proposed local sensitivity to fix one of any two adjacent datasets to the actual dataset being queried and take into account all its neighbors. But it does not meet the requirements of differential privacy. Smooth sensitivity~\cite{Nissim2007SmoothSA} is the tightest smooth upper bound of local sensitivity which can prevent the privacy leakage caused by local sensitivity, but its computation complexity is NP-hard. Elastic sensitivity(ES)~\cite{Johnson2017TowardsPD} and residual sensitivity(RS)~\cite{Dong2021ResidualSF} are both based on the idea of finding the smooth upper bound of the local sensitivity. ES is computed based on the  maximum frequency of the join values in each relation, and it can be computed fast but introduces noise orders-of-magnitude higher than the query answer.  RS is computed based on some residual queries of a multi-join query, and the maximum boundary of a residual query measures the influence of adding a tuple to a certain relation by computing upper bound of joining such a tuple with other private relations.

Approximate query processing (AQP) is a technique that can be used to estimate join sizes efficiently~\cite{Chaudhuri2017ApproximateQP}.
Sampling of the join result is one of the AQP methods with high performance. Ripple join~\cite{Haas1999RippleJF} and wander join ~\cite{Li2016WanderJO} are online aggregations which can be used for the queries with join operators. Zhao et al~\cite{Zhao2018RandomSO}. revisited this issue to integrate the previous approaches into a universal framework which contains two main phrases, calculating the upper bound of join size weight and then sampling from joins. 
Sketches are probabilistic data structures used for stream summarization tasks, and masses of sketches such as AGMS~\cite{DBLP:conf/sigmod/DobraGGR02}, Count-sketches~\cite{charikar2002finding}, and Count-Min sketches~\cite{cormode2005improved} are proposed for frequency estimation, heavy hitter mining, and join size estimation, etc. AGMS sketch is designed for self-join estimation, which can also be used to estimate the join size of multi-join queries including condition filters~\cite{DBLP:journals/pvldb/VengerovMZC15}.  

Though a variety of definitions for sensitivity is proposed for join queries, they all have some shortcomings.
Global sensitivity is unbounded for multi-join queries. Local sensitivity does not satisfy the differential privacy. The computation of smooth sensitivity is NP-hard. ES is easy to compute but with poor accuracy. RS is more accurate but more time-consuming. No existing works have make efforts to adopting AQP methods for sensitivity estimation. However, AQP methods such as sampling and sketches cannot be simply adopted to reduce the cost. The reason is that the sensitivity of a multi-join queries depends on many statistics as shown in RS, and individually estimating each statistics is still costly. Therefore, we propose a sampling-based sensitivity estimation method to focus on estimating the statistics have contributes to the sensitivity. We also propose a sketch-based method which constructs sketches for each relation offline and efficiently estimates the sensitivity online.


\section{Preliminaries}
In this section, we introduce some definitions about differential privacy and different kind of sensitivity. The symbols used in this section is shown in Table~\ref{table:notation-for-priliminary}.

\begin{table}[!h]
\caption{Table of notations.}
\label{table:notation-for-priliminary}
\centering
\footnotesize
\begin{tabular}{ll}
 \toprule
 Notation & Meaning \\
 \midrule
$\epsilon$ & Privacy budget \\
$\delta$ &  The probability pure differential privacy fails to hold.\\
$LS_q^k(I)$ & Local sensitivity of $q$ on a database at distance $k$ from $I$.\\
$\widetilde{LS}_q^k(I)$ & The upper bound of $LS_q^k(I)$ computed by ES.\\
$\hat{LS}_q^k(I)$ & The upper bound of $LS_q^k(I)$ computed by RS.\\
$mf(A)$ & The frequency of the most frequent value of attribute $A$. \\
$T_E(I)$ & The maximum boundary of residual query $q_E$. \\
\bottomrule
\end{tabular}
\end{table}

\subsection{Differential Privacy}
Differential Privacy(DP)~\cite{Dwork2006DifferentialP} was originally proposed by Dwork in 2006 for the privacy leakage problem of statistical databases, which ensures that any individual in the dataset or not has little effect on the final released query results.

\begin{definition}
\textbf{$(\epsilon,\delta)$-Differential Privacy.}
  A randomized privacy algorithm $A$ satisfies $\epsilon$-DP if it for any pairs of input datasets $I$ and $I'$ satisfying $d(I,I')=1$, for all sets $S$ of possible outputs.
  \begin{equation}
    \Pr[A(I) \in S] \le exp(\epsilon)\cdot\Pr[A(I') \in S]+\delta
  \end{equation}
  The parameter $\epsilon$ set artificially is also known as privacy budget, which indicates the degree of privacy protection. The smaller the value, the higher the degree of protection. When $delta=0$, we say the algorithm satisfy $\epsilon$-Differential Privacy.
\end{definition}

\begin{theorem}
\textbf{Laplace Mechanism}~\cite{Dwork2006CalibratingNT}
 Given a dataset $I$, function $f$:$I \rightarrow R^d$. So the random algorithm $A:A(I)=f(I)+Y$ provides $\epsilon$-differential privacy protection, and $Y \sim Lap(\bigtriangleup f/\epsilon)$ presents the randomized noise, which follows Laplace distribution with scale parameter $\bigtriangleup f/\epsilon$.
 Apparently, the parameter $\bigtriangleup f$ known as the ``sensitivity'' plays an important role in calculating noise.
\end{theorem}

\subsection{Sensitivity}
Sensitivity is used to measure the maximum modification to the query result when inserting or deleting a tuple in the result. In this section, we introduce five different definitions of sensitivity.

Global sensitivity is defined as follows, which calculates the maximum difference of the query result on two databases whose distance is one, i.e., $d(I,I')=1$. The global sensitivity only relates to $q$. For example, the global sensitivity of a simple COUNT query is 1, no matter querying on which database. But the join query sensitivity is unbounded large according to this definition.

\begin{definition}
\textbf{Global Sensitivity.~\cite{Dwork2006CalibratingNT}}
  For $q:D^n\rightarrow R^d$ and all $I,I'\in D^n$, the global sensitivity of $q$ is
  \begin{equation}
    GS_q=\max\limits_{I,I':d(I,I')=1}||q(I)-q(I')||
  \end{equation}
\end{definition}

Local sensitivity is defined in the similar way, and the only difference is that it is computed for fixed database.
\begin{definition}
 \textbf{Local Sensitivity.~\cite{Nissim2007SmoothSA}}
  For $q:D^n\rightarrow R^d$ and $I'\in D^n$, the local sensitivity of $q$ for $I$ is
  \begin{equation}
    LS_q(I)=\max\limits_{I':d(I,I')=1}||q(I)-q(I')||
  \end{equation}
\end{definition}

Local sensitivity introduces less noise, but releasing the query result according to local sensitivity does not satisfy the definition of differential privacy. To solve this problem, smooth sensitivity is proposed to compute the smooth upper bound on the local sensitivity.
\begin{definition}
  \textbf{Smooth Sensitivity.~\cite{Nissim2007SmoothSA}}
  The smooth sensitivity is defined based on the generalization of local sensitivity of $f$ at distance $k$,
  \begin{equation}
    LS_q^k(I)=\max\limits_{I'\in D^n:d(I,I')=k}LS_q(I')
  \end{equation}
  the smooth sensitivity of $q$  for $I$ is
  \begin{equation}
    SS_q(I)=\max\limits_{0\le k\le n}e^{-\beta k} LS_q^k(I), \beta=\frac{\epsilon}{2\ln 2/\delta}
  \end{equation}
\end{definition}

Computing the smooth sensitivity takes exponential time, which is very expensive to the query on big data. To reduce the cost, elastic sensitivity and residual sensitivity are proposed to compute the sensitivity for multi-way joins, both of which are smooth upper bounds of local sensitivity. Nissim et al.~\cite{Nissim2007SmoothSA} proved that any smooth upper bound of local sensitivity can preserve the privacy according to differential privacy.

\begin{definition}
  \textbf{Elastic Sensitivity.~\cite{Johnson2017TowardsPD}}
 The elastic sensitivity of a database $I$ is defined based on the smooth upper bound of local sensitivity as follows.
 \begin{equation}
    ES_q(I)=\max\limits_{0\le k\le n}e^{-\beta k} \widetilde{LS}_q^k(I), \beta=\frac{\epsilon}{2\ln 2/\delta}
  \end{equation}
  where $\widetilde{LS}_q^k(I)$ is an upper bound of local sensitivity at distance $k$,
  \begin{equation}
  \begin{split}
    \widetilde{LS}_q^k(I)=\max\limits_{i\in P}(\prod_{j\in P-\{i\}}(mf(x_j\cap x_{p(j,i)},I_j)+k)\\\cdot
    \prod_{j\in [n]-P}mf(x_j\cap x_{p(j,i)},I_j))
    \end{split}
  \end{equation}
   $mf(x,I_j)$ is the maximum frequency on attribute $x$ in $I_j$, and $P$ is the private attribute set.
\end{definition}
Elastic sensitivity can be computed based on the maximum frequencies of the join attributes. But it computes the maximum frequency of the multi-way join based on the assumption that all the  most frequent join attribute values  can be joined with each other. So, elastic sensitivity involves much error.

\begin{definition}\label{def:RS}
  \textbf{Residual Sensitivity.~\cite{Dong2021ResidualSF}}
  The residual sensitivity of a database $I$ is also defined based on the smooth upper bound of local sensitivity at distance $k$,
  \begin{equation}\label{equ:RS}
    RS_q(I)=\max\limits_{0\le k\le n}e^{-\beta k}\min(\hat{GS_q}, \hat{LS}_q^k(I)), \beta=\frac{\epsilon}{2\ln 2/\delta}
  \end{equation}
  where $\hat{LS}_q^k(I))$ is an upper bound of local sensitivity at distance $k$,
  \begin{equation}
    \hat{LS}_q^k(I)=\max\limits_{s\in S^k}\max\limits_{i\in P}\hat{T}_{[n]-\{i\},s}(I)
  \end{equation}
  \begin{equation}\label{equ:hatTE}
  \hat{T}_{E,s}(I)=\sum_{E'\subseteq E}\left(T_{E-E'}(I)\prod_{i\in E'}s_i\right)
  \end{equation}
  $\hat{T}_{E,s}(I)$ computes the maximum boundary for the $I':d(I,I')=k$, $s$ is one way of partitioning $k$ into different relations in $E$.
\end{definition}

We use the following example to demonstrate the computation of residual sensitivity.
\begin{example}
  Consider the query $q=R_1\Join R_2\Join R_3$ in Fig 1, we suppose $R_1$,$R_2$, and $R_3$ are all private relations. According to the residual sensitivity definition,
  \begin{equation}
    \hat{LS}_q^k(I)=\max\limits_{s\in S^k}\max\{\hat{T}_{\{1,2\},s}(I),\hat{T}_{\{2,3\},s}(I),\hat{T}_{\{1,3\},s}(I)\}
  \end{equation}
  where $s={s_1,s_2,s_3}$, and $s_1+s_2+s_3=k$.
  $\hat{T}_{\{1,2\},s}(I)$  is the upper bound of the following query:

  SELECT MAX(count) from (SELECT $\hat{R2}$.C, COUNT(*) from $\hat{R1}$,$\hat{R2}$ where $\hat{R1}$.B=$\hat{R2.B}$ GROUP BY $\hat{R2}$.C) as T;

  Here $\hat{R1}$ is the relation after inserting $s_1$ tuples, and $\hat{R2}$ is the relation after inserting $s_2$ tuples.

  According to equation~\ref{equ:hatTE},
  \begin{equation}
   \hat{T}_{\{1,2\},s}(I)=T_{\{1,2\}}(I)+s_1\cdot T_{\{2\}}(I)+s_2\cdot T_{\{1\}}(I)+s_1\cdot s_2
   \end{equation}
We can calculate $\hat{T}_{\{2,3\},s}(I)$, and  $\hat{T}_{\{1,3\},s}(I)$ in the same way. At last, the residual sensitivity can be computed according to equation~\ref{equ:RS}.
\end{example}

As shown in the above example, residual sensitivity requires computing each $T_E$, where $E\subseteq[n]-\{i\}$ for each private relation $R_i\in\{R_1,...,R_n\}$. Each $T_E$ can be computed by an aggregation query in form of the query in the example, and computation is still costly when it contains join operators.

\section{Sensitivity Estimation for Join Queries}

In this section, we propose two sensitivity estimation methods based on sampling and sketch, respectively.
The notations in this section is shown in Table~\ref{table:notation-for-SE}.
\begin{table}[!h]
\caption{Notations for sensitivity estimation.}
\label{table:notation-for-SE}
\centering
\footnotesize
\begin{tabular}{ll}
 \toprule
 Notation & Meaning \\
 \midrule
 $q_{E}$ & A residual query on a subset $E$ of a multi-join query $q$.\\
$m_{E,i}$ & The sample size for the $i$th group of residual query $q_E$.\\
$\tau_{E,i}$ & Half-width of the confidence interval for the $i$th group of $q_E$.\\
$g$ & The number of groups of a residual query.\\
$J$ & Join size\\
$\eta$ & The probability that the confidence interval fails to hold\\
$sk(R)$ & The AGMS sketch of a relation $R$\\
\bottomrule
\end{tabular}
\end{table}

\subsection{Sampling-based Sensitivity Estimation}
Calculating the residual sensitivity for a multi-way join query is expensive, since it computes group-by aggregations for a list of residual queries. A basic idea to reduce the cost is to use sampling methods to estimating the the results of each residual queries in form of $Q_1$ in example 1.  Inspired by ``Wander-Join''\cite{DBLP:conf/sigmod/0001WYZ16}, we introduce a method to fast estimate such a residual query based on random walk in section~\ref{Est-for-one-RS-query}. As a residual query result is only relevant to the largest group of the join result, we focus on the join paths for the largest group. To further reduce the cost, we proposed a method to estimate all the residual queries using a set of join paths, which is described in section~\ref{sec:Improved-Sampling-SE}.
\subsubsection{Estimation for one residual query}\label{Est-for-one-RS-query}
Calculating the true result of the maximum boundary of each residual query is costly. To this end, we present a sampling-based sensitivity estimation method (Sampling-SE).
\begin{algorithm}
\caption{Sampling-SE}\label{Algorithm:Sampling-SE}
{\bf Input:} Multi-Join query $q$ \\
{\bf Output:} The sensitivity of $q$
\begin{algorithmic}[1]
\For {Each residual query $q_E$ of $q$}
    \State $T_E \leftarrow RQE(q_E)$
\EndFor
\State Compute the sensitivity based on $T_E$ for each $q_E$ according to Definition~\ref{def:RS}.
\end{algorithmic}
\end{algorithm}
Inspired by the ``Wander-Join''\cite{DBLP:conf/sigmod/0001WYZ16}, we start the algorithm by sampling the join path of each group in a round-robin fashion. In order to make sure all the groups are well estimated, ``Wander-Join'' iteratively selects the group that has the largest confidence interval to start the next random walk. Unlike this, we do not care about all the groups, but only about the largest group. To address this, we adopt the idea similar to ``iFOCUS''~\cite{DBLP:journals/pvldb/KimBPIMR15} to sample more for the groups we care about. The difference is that, ``iFOCUS'' keeps removing the groups having no overlapping confidence interval with others for ordering guarantee, while we keep removing the group whose confidence interval has no overlap with the largest group.

The pseudo-code of the Sampling-SE algorithm is shown in Algorithm~\ref{Algorithm:Sampling-SE}. As computing the $T_E$ for each residual query of a multi-join query is costly, we use a sampling-based method $RQE$ to estimate the maximum boundary $T_E$ of each residual query $q_E$.

\begin{algorithm}
\caption{RQE}\label{Algorithm:Est_One_ResidualQuery}
{\bf Input:}
Residual query $q_E=\Join_{i\in E}R_i$, Error bound $\tau_0$  \\
{\bf Output:}
The estimation $T_E$
\begin{algorithmic}[1]
\State Initial the count estimation $C_1, C_2,...,C_g$ for each distinct values  $v_1, v_2,...,v_g$ with $m$ random walks. $G=\{1,2,...,g\}$
\State Initial the join size $J$ and error bound $\tau_{J}$ for $q_E$ according to the random walks in step 1.
\State $n\leftarrow m\cdot g$
\While {$\tau>\tau_0$}
    \State $m\leftarrow m+1$
    \For {each $i\in G$}
        \State $n\leftarrow n+1$
        \State Conduct a random walk $p$ starting from $t(v_i)$.
        \State $C_i$, $\tau_i\leftarrow$ Estimate($p$, $t(v_i)\Join q_E$, $m$, $C$) 
        \State $J$, $\tau_J\leftarrow$ Estimate($p$, $q_E$, $n$, $J$)
        \State $\tau\leftarrow\tau_i\cdot(J+\tau_J)$

    \EndFor
    \For {each $i\in G$}
        \If {$C_i+\tau < \max\limits_{j\in G}(C_j-\tau$)}
            \State $G\leftarrow G -\{i\} $
        \EndIf
    \EndFor
\EndWhile
\State \textbf{return: $T_E=max_{i\in G}(C_i+\tau)$}
\end{algorithmic}
\end{algorithm}

\begin{algorithm}
\caption{Estimate($p$, $q$, $m$, $C$)}\label{Algorithm:updateCI}
\begin{algorithmic}[1]
\State $x\leftarrow$ Estimate $|q|$ with path $p$ according to WanderJoin.
\State $C \leftarrow \frac{m-1}{m}C +\frac{1}{m}x $
\State $\tau=\sqrt{\frac{2\log\log(m)+\log((g+1)\pi^2/6\eta)}{2m}}$
\State \textbf{Return: $C$, $\tau$.}
\end{algorithmic}
\end{algorithm}

Details of the residual query estimation(RQE) method is shown in Algorithm~\ref{Algorithm:Est_One_ResidualQuery}. The algorithm first conducts $m$ join paths for each group according to the random walk, and we initialize the estimation $C_1,C_2,...,C_g$ for each group in $G=\{1,2,...,g\}$ (line 1). We then estimate the join size $J$ and half-width of the confidence interval $\tau_J$ for the query $q_E$. The main part of this algorithm (line 3-18) iteratively increases the sample size for each candidate large group, whose confidence interval overlaps with the current largest group. For the candidate large groups in $G$, the algorithm adds random walk path to update the estimations (line 6-12). The algorithm $Estimate$ computes the new estimate $C_i$ and half-width of confidence interval $\tau_i$ for the $i$th group according to the join path $p$ (line 9), and then it estimates the join size and the confidence interval for $J$ in the same way (line 10). For the small groups whose confidence intervals have no overlap with the current largest group, we remove them from $G$ (line 13-17). The algorithm stops when the half-width of the confidence interval $\tau$ is below $\tau_0$, meaning that the estimation is accurate enough.

Algorithm~\ref{Algorithm:updateCI} estimates the result size $C$ of a query $q$ based on the join path $p$, and it computes the half-width of confidence interval $\tau$ according to Hoeffding Inequality~\cite{0Probability}.


We use two steps to prove that the output of the  algorithm is a sufficiently accurate estimate for each $T_E$: (1) the algorithm do not miss the largest group, and (2) the probability that true largest group size is larger than the output of the algorithm is smaller than $\eta$.

\textbf{First step.} we use the following Theorem~\ref{no-miss} to prove that the algorithm do not miss the largest group.
\begin{theorem}\label{no-miss}
  If for each group $i$, we have $|C_i-\mu_i|\le \tau$ for every $1\le m\le N$, then largest group $j\in G$ at termination time.
\end{theorem}
\begin{proof}
Assuming the largest group $j\notin G_{terminate}$, then there exists a group $k$ whose lower bound is higher than the upper bound of group $j$, i.e, $C_j+\tau<C_k-\tau$ according to Algorithm~\ref{Algorithm:Est_One_ResidualQuery}(line 14). Since $|C_i-\mu_i|\le \tau$ holds for every $1\le m\le N$,
\begin{equation}
  \mu_j \le C_j+\tau<C_k-\tau\le\mu_k
\end{equation}
As $\mu_j<\mu_k$, $j$ is not the largest group, which contradicts the assumption, so the assumption is not true.
\end{proof}

\textbf{Second step.} We used the Theorem~\ref{failure limit} to prove that the probability that the  true result is larger than the output of the algorithm is limited.
\begin{lemma}
  \textbf{Hoeffding Inequality~\cite{0Probability}.} Let $Y=y_1$, $y_2$,...,$y_N$ be a set of $N$ values in [0,1] with average value $\frac{1}{N}\sum_{i=1}^{N} y_i =\mu$. Let $X_1$,...,$X_m$ be a sequence of random variables drawn from $Y$ without replacement. For every $1\le g\le N$ and $\epsilon>0$,
  \begin{equation}\label{EQU:Hoef-Serf-Ineq}
    \Pr[\max\limits_{g\le m\le N-1}(\frac{\sum_{i=1}^{m}X_i}{m}-\mu)\ge \tau]\le \exp(-2g\tau^2)
  \end{equation}
\end{lemma}

Suppose $J$ is the total join size of $\{\Join_{i\in E}R_i\}$, we divide the join size for each group $C_i$ by $J$ to make each $\frac{C_i}{J}\in [0,1]$. Thus, we can use the above inequality to get the error bound for each group.

\begin{theorem}\label{failure limit}
For all the groups in $G$ the join size, for all the rounds in Algorithm 1, we have $\Pr[\exists i, m, 1\le i\le g,1\le m\le m_i:(est_{i,m}-\mu_i)\ge J_{up}\cdot\tau]\le \eta$, where $\tau =\sqrt{\frac{2\log\log(m)+\log((g+1)\pi^2/6\eta)}{2m}}$, and $J_{up}=J_{est}+\sqrt{\frac{2\log\log(n)+\log((g+1)\pi^2/6\eta)}{2n}}$. Here, $m$ is the number of samples to estimate each group size, $n$ is the number of samples to estimate the total join size.
\end{theorem}

\begin{proof}
  We prove the theorem in similar way with the theorem 3.2 in IFOCUS~\cite{DBLP:journals/pvldb/KimBPIMR15}, and the difference is that we only need to prove that the probability that the estimate beyond the upper bound of the confidence interval is limited. We use the above lemma to compute the upper bound for each group as follows.
  \begin{equation}\label{equ:epsilonm}
    \tau_i=\sqrt{\frac{(2\log\log(m)+\log(\pi^2/6\eta_i))}{2m}}
  \end{equation}
  Then, $\Pr[\exists m,1\le m\le N:(\frac{\sum_{i=1}^{m}X_i}{m}-\mu)>\tau]\le \eta_i$.
  \begin{equation}
  \begin{split}
   & \Pr[\exists m,1\le m\le N:(\frac{\sum_{i=1}^{m}X_i}{m}-\mu)>\tau_m]\\
   & \le\sum_{r=1}\Pr[\exists m,\kappa^{r-1}\le m \le \kappa^{r}:(\frac{\sum_{i=1}^{m}X_i}{m}-\mu)>\tau_m]\\
   & \le\sum_{r=1}\Pr[\max\limits_{\kappa^{r-1}\le m \le N-1}(\frac{\sum_{i=1}^{m}X_i}{m}-\mu)>\tau_{\kappa^r}]
   \end{split}
  \end{equation}
  By the Lemma 1,
  \begin{equation}\Pr[\max\limits_{\kappa^{r-1}\le m \le N-1}(\frac{\sum_{i=1}^{m}X_i}{m}-\mu)>\tau_{\kappa^r}]\le \frac{6\eta}{\pi^2 r^2}.
  \end{equation}
As $\sum_{r\ge 1}\frac{1}{r^2}=\frac{\pi^2}{6}$,
  \begin{equation}
    \sum_{r=1}\Pr[\max\limits_{\kappa^{r-1}\le m \le N-1}(\frac{\sum_{i=1}^{m}X_i}{m}-\mu)>\tau_{\kappa^r}]
    \le \eta
  \end{equation}

  The equation~\ref{equ:epsilonm} computes the half-width $\tau_i$ of the confidence interval for the estimation of $\frac{C_i}{N}$. We can easily multiply $\tau_i$ with the total join size $N=J$ to calculate the half-width of the confidence interval for $C_i$. The join size $J$ is unknown in advance, but it can also be estimated according to the random walks we pick in each round. Suppose we get $n$ paths to estimate $J$,
  \begin{equation}\label{equ:epsilonn}
    \tau_J=\sqrt{\frac{(2\log\log(n)+\log(\pi^2/6\eta))}{2n}}
  \end{equation}
  Then, $\Pr[\exists n,1\le n\le N:|\frac{\sum_{i=1}^{n}X_i}{n}-\mu|>\tau_n]\le \eta$.

Regarding $J_{up}$ as the upper bound for a big group including all the join results, we get $g+1$ groups. Suppose $\eta_i=\eta_j$ for each pair $i,j\in \{0,1,...,g+1\}$, we get $\eta_i=\eta/(g+1)$. Setting $\eta$ in equation~\ref{equ:epsilonm} and equation~\ref{equ:epsilonn} as $\eta/(g+1)$, we get the $\tau$ in line 11 of algorithm~\ref{Algorithm:Est_One_ResidualQuery}.
\end{proof}

As many groups are removed once the upper bound of their estimates are below the lower bound of the largest group, the sample complexity for each group is different. We  use the following theorem to prove the sample complexity of our algorithm.

\begin{theorem}
  \textbf{Sample Complexity.} With probability at least $1-\eta$, the Algorithm~\ref{Algorithm:Est_One_ResidualQuery} outputs the upper bound for the largest group, and draws
  $O\left( J_{up}^2\sum_{i=1}^{g}\frac{\log(\frac{g}{\eta})+\log\log(\frac{1}{\alpha_i})}{\alpha_i^2} \right)$ samples. Here, $\alpha_i=\max\{\frac{|\mu_{max}-\mu_i|}{4},\tau_0\}$
\end{theorem}

\begin{proof}
  We first prove that the sampling stop for each group $i$ once the $\tau\le \max\{\frac{|\mu_{max}-\mu_i|}{4},\tau_0\}$.

  If group $i$ is removed from $G$ before $\tau$ reaches $\tau_0$, i.e., $\tau>\tau_0$, then $C_i+\tau<C_{max}-\tau$. Since the confidence interval always contains the true result, $\mu_{max}\in [C_{max}-\tau,C_{max}+\tau]$, and $\mu_i\in [C_i-\tau,C_i+\tau]$.  To make sure  $C_i+\tau<C_{max}-\tau$ holds for the worst case, we get $\tau\le \frac{|\mu_{max}-\mu_i|}{4}$.

  If group $i$ is not removed from $G$ until $\tau$ reaches $\tau_0$, then $\tau=\tau_0$.

  Thus, $\alpha_i=\max\{\frac{|\mu_{max}-\mu_i|}{4},\tau_0\}$ is the half-width of the confidence interval for group $i$ when stop adding samples. Let $\epsilon=\alpha_i$, we get
  \begin{equation}
    m_i=O\left(J_{up}^2 \frac{\log(\frac{g}{\eta})+\log\log(\frac{1}{\alpha_i})}{\alpha_i^2}\right)
  \end{equation}
  The sample complexity for all the groups is
  \begin{equation}
  O\left( J_{up}^2\sum_{i=1}^{g}\frac{\log(\frac{g}{\eta})+\log\log(\frac{1}{\alpha_i})}{\alpha_i^2} \right).
   \end{equation}
\end{proof}

We can infer from the theorem above that the sample size complexity is closely linked to the distance between the size of the largest group and that of the other groups. The greater the distance, the fewer samples are required.

\subsubsection{Improved Sampling-based Sensitivity Estimation}\label{sec:Improved-Sampling-SE}

\begin{figure}[htbp]
\centering
\includegraphics[scale=0.3]{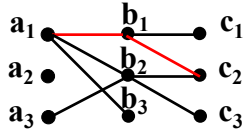}
\caption{Estimations of values on one join path.}
\label{Fig:RQS_example}
\end{figure}

In the previous section, we introduce the estimating algorithm for each residual query of a multi-way join query. However, to compute the residual sensitivity requires calculating all the residual results in $\{q_E|E\subseteq{nr-\{i\}},i\in P\}$. We can simply use the algorithm~\ref{Algorithm:Est_One_ResidualQuery} to estimate each residual query, but it involves redundant samples.
As shown in figure~\ref{Fig:RQS_example}, with a join path $a_1\rightarrow b_2\rightarrow c_2$, we can estimate   $|t(a_1)\Join R_1\Join R_2\Join R_3|=\frac{1}{\frac{1}{3}\cdot \frac{1}{2}}=6$ and $|t(b_1)\Join R_2\Join R_3|=\frac{1}{\frac{1}{2}}=2$.
Therefore, we propose an algorithm to reduce the sample complexity by leveraging each path to estimate the candidate of all the values on the path.

\begin{algorithm}
\caption{Improved-RQE}\label{Algorithm:Improved-RQE}
{\bf Input:}
Multi-way Query $q$,
Residual queries $RQS=\{q_E|E\subseteq{[n]-\{i\}},i\in P\}$,
Terminal error bound $\tau_0$  \\
{\bf Output:}
The estimations  $\{T_E|q_E\in RQS\}$
\begin{algorithmic}[1]
\State $ARQS\leftarrow RQS$
\While {$ARQS\ne\emptyset$}
    \State $q_E\leftarrow \arg\max\limits_{q_E\in ARQS}|E|$
        \For {each group $i\in G_E$}
            \State Conduct a random walk $p$ starting from $t(v_i)$.
            \For {each $E'\subseteq E$ on $p$}
                \State $m_{E',i}\leftarrow m_{E',i}+1$
                \State $n_{E'}\leftarrow n_{E'}+1$
                \State $C_{E',i}$, $\tau_{E',i}\leftarrow$ Estimate($p$, $t(v_i)\Join q_E'$, $m$, $C_{E',i}$) 
                \State $J_{E'}$, $\tau_{J_{E'}}\leftarrow$ Estimate($p$, $q_{E'}$, $n$, $J$)
                \State $\tau_{E',i}\leftarrow\tau_{E',i}\cdot(J_{E'}+\tau_{J_{E'}})$
            \EndFor
        \EndFor
    \For {each $q_E\in ARQS$}
        \For {each $i\in G_E$}
            \If {$C_{E,i}+\tau_{E,i} < \max\limits_{j\in G_E}( C_{E,j}-\tau_{E,j}$)}
                \State $G_E\leftarrow G_E -\{i\} $
            \EndIf
        \EndFor
        \If {$\tau_{E,i}<=\tau_0$ for each group $i\in{G_{E}}$ }
                \State $ARQS\leftarrow ARQS -\{q_E\}$
        \EndIf
    \EndFor
\EndWhile
\State \textbf{return:$\{T_E=max_{i\in G_E}(C_{E,i}+\tau_{E,i})|q_E\in RQS\}$ }
\end{algorithmic}
\end{algorithm}

The pseudo-code is shown in Algorithm~\ref{Algorithm:Improved-RQE}.  As the basic idea is to estimate the join size of each value on the path to reduce the sample complexity, we start the algorithm to draw samples for the residual query with the most relations. Because a long path can provide estimations for more values.
The algorithm then draws new samples for the groups of $q_E$ to update the estimations and confidence intervals for each $E'\subseteq E$ (line 4-13). The algorithm iteratively removes the group from the group set $G_E$ of each $q_E\in ARQS$ if it is not the candidate largest group (line 15-19), and removes the $q_E$ from $ARQS$ once all the remaining groups of $q_E$ is well estimated (line 20-22). The algorithm stops once the active residual query set ($ARQS$) is empty.

In this algorithm, one join path can be used to estimate the join size of each value on the path. Thus, the sample complexity can be reduced.

\subsection{Sketches-based Sensitivity Estimation}
Sketches are useful data stream summaries widely used for frequency estimation, heavy hitter finding, and join size estimation. In this section, we propose a sensitivity estimation method based on AGMS sketch.
\subsubsection{Sketches based Multi-join Size Estimation}
The basic idea of AGMS is mapping the values $v_1$, $v_2$, ..., $v_{|dom(A)|}$ of a join attribute $A$ in a relation $R$ into four-wise random variables $\xi(v_1)$, $\xi(v_2)$, ..., $\xi(v_{|dom(A)|})$, where each $\xi(v_i) \in \{-1,+1\}$ and $\Pr[\xi(v_i)=+1]=\Pr[\xi(v_i)=-1]=1$. The AGMS sketch for a relation $R$ is
\begin{equation}
  sk(R)=\sum_{i\in dom(A)}f(i)\xi(v_i),
\end{equation}
where $f(i)$ is the frequency of $v_i$. The $sk(R)$ can be computed with one-pass scanning of $R$.
The product of two sketches $sk(R_1)$ and $sk(R_2)$ for two relations $R_1$ and $R_2$ is an unbiased estimate of the join size of $R_1\Join R_2$: $\mathbb{E}[sk(R_1)\cdot sk(R_2)]=|R_1\Join R_2|$.

Each relation in a multi-join query may contain multiple join attributes.
The AGMS can be used to estimate multi-join size by defining a distinct random family $\xi_1$, $\xi_2$, ..., $\xi_n$  for each equi-join attribute pairs. The sketch for each relation $R$ can be written as:
\begin{equation}
sk(R)=\sum_{t\in R}\prod\limits_{i\in JA} \xi_i(t[i]),
\end{equation}
 where $JA$ is the set of all the join-pair attributes, and $t[i]$ is the value of attribute $i$ of tuple $t$. We use the following example to show how to compute the sketches for the relations in Figure~\ref{Fig:intro_example}.
\begin{example}
  Consider the example in Fig 1, we define two families of four-wise independent random variables for the join attributes $B$ and $C$ as $\xi_1$ and $\xi_2$. Three separate sketches are constructed for $R_1(A,B)$, $R_2(B,C)$, and $R_3(C,D)$ as

$sk(R_1)=\sum_{t\in R_1} \xi_1(t[B])$,

$sk(R_2)=\sum_{t\in R_2} \xi_1(t[B])\cdot  \xi_2(t[C])$,

$sk(R_3)=\sum_{t\in R_3} \xi_2(t[C])$.

The value of $X=sk(R_1)\cdot sk(R_2)\cdot sk(R_3)$ gives an unbiased estimate of for $R_1\Join R_2\Join R3$.
\end{example}
Though one estimate is not sufficiently accurate, boosting technique can further improve the accuracy by conducting averaging and median-selection on several independent estimates. The final boosted estimate is the median of $s_2$ variables $Y_1$,..., $Y_{s2}$, where each $Y_i$ is the average of $s_1$ independent estimates $X_1$,..., $X_{s1}$. To simplify the expression, we only use the mean of $s1$ independent estimates to denote the join size estimation based on AGMS sketch in the following parts.

\subsubsection{Sketching Sensitivity for Multi-join Queries}
We define the \textit{sketching sensitivity} of a multi-join query in terms of $sk^{(k)}(R_i)$, the sketch of the relation $R_i$ at distance $k$ from the database. And we build a connection between local sensitivity and sketching sensitivity.

We first consider estimating the  local sensitivity based on sketches as follows.
\begin{theorem}
  Local sensitivity can be computed as:
  \begin{equation}\label{equa:Sketch-Local-sensitivity}
  LS_{q}(I)=\max\limits_{i\in P}\max\limits_{v\in dom(JA(R_i))}(v\Join_{j\in [r]-\{i\}} R_{j,I})
\end{equation}
where $R_{j,I}$ is the $j$th relation in instance $I$.
\end{theorem}

According to the following theorem proved in reference~\cite{DBLP:conf/sigmod/DobraGGR02}, the relative error of the join size estimate based sketches can be limited.

\begin{theorem}
  Let Q be an acyclic, multi-join query over relations $R_1$,...$R_r$, such that $Count(Q)\ge L$ and Self-Join($sk_k$)$\le U_k$. Then, using a sketch of size $O(\frac{2^{2n}(\prod_{k=1}^{2}U_k)log(1/\eta)}{L^2 \tau^2} \sum_{j=1}^{n}\log|dom(A_j)|)$, it is possible to approximate $Count(Q)$ so that the relative error of the estimate is at most $\tau$ with probability at least $1-\eta$.
\end{theorem}

We compute the upper bound of join size by dividing the estimate by $(1-\tau)$, thus, with the probability at least $1-\eta$, the true join size $J<\frac{est}{1-\tau}$.

Each $v\Join_{j\in [n]-\{i\}} R_{j,I}$ in equation~\ref{equa:Sketch-Local-sensitivity} can be estimated based on the sketch of each relation:
\begin{equation}\label{equ:SketchLS-upperbound}
\begin{split}
 &(v\Join_{j\in [n]-\{i\}} R_j)\\
 &\le\frac{\mathop{mean}\limits_{s\in[1,s1]} \left( \prod\limits_{l\in JA(R_{i,I}) }\xi_{l,s}(v)\cdot \prod_{j\in [n]-\{i\}}sk_s(R_{j,I})\right)}{1-\tau}\\
 &\le\frac{\mathop{mean}\limits_{s\in[1,s1]} \left|\prod_{j\in [n]-\{i\}}sk_s(R_{j,I})\right|}{1-\tau}
\end{split}
\end{equation}
with the probability at least $1-\eta$. The second inequality of equation~\ref{equ:SketchLS-upperbound} holds because  $\prod\limits_{l\in JA(R_{i,I}) }\xi_{l,s}(v)\in\{-1,1\}$.

Since the smooth upper bound of local sensitivity is defined based on the local sensitivity at distance $k$, we also consider to estimate that according to sketches.
\begin{theorem}
  Suppose $\mathcal{I}^k=\{I':d(I,I')=k\}$ is the set of instance having distance $k$ to $I$, local sensitivity at distance $k$ can be computed as:
    \begin{equation}
      LS_{q}^{(k)}(I)=\max\limits_{I'\in \mathcal{I}^k}\max\limits_{i\in P}\max\limits_{v\in dom(JA(R_i))}(v\Join_{j\in [r]-\{i\}} R_{j,I'})
    \end{equation}
\end{theorem}

  Let $S^k=\{(k_1,k_2,...,k_r)|\sum k_i=k,  k_{x\notin P}=0)\}$ be the set of all the partitions of $k$ tuples, the $LS_{q}^{(k)}(I)$ can be rewrite as follows.
  \begin{equation}
  \!LS_{q}^{(k)}(I)=\!\!\!\!\!\max\limits_{(k_1,k_2,...k_r)\in S^k}\!\max\limits_{i\in P}\!\max\limits_{v\in dom(JA(R_i))}\!(v\Join_{j\in [r]-\{i\}}\!R_j^{(k_j)})
\end{equation}
Each $v\Join_{j\in [n]-\{i\}}\!R_j^{(k_j)}$ can be estimated based on the sketch of each relation:
\begin{equation}\label{equa:Sketch-Local-sensitivity-distance-k}
\begin{split}
&(v\Join_{j\in [n]-\{i\}} R_j)\\
&<=\frac{\mathop{mean}\limits_{s\in[1,s1]} \left( \prod\limits_{l\in JA(R_i) }\xi_{l,s}(v)\cdot \prod_{j\in [n]-\{i\}}sk_s^{(k_j)}(R_j)\right)}{1-\tau}\\
&<=\frac{\mathop{mean}\limits_{s\in[1,s1]} \left|\prod_{j\in [n]-\{i\}}sk_s^{(k_j)}(R_j)\right|}{1-\tau}
\end{split}
\end{equation}
Therefore,
\begin{equation}\label{equ:SketchLS-upperbound-distance-k}
  LS_{q}^{(k)}(I)\le\max\limits_{I'\in \mathcal{I}^k}\max\limits_{i\in P}\frac{\mathop{mean}\limits_{s\in[1,s1]} \left|\prod_{j\in [n]-\{i\}}sk_s^{(k_j)}(R_j)\right|}{1-\tau}
\end{equation}

We can regard the right-side of the above equation as the upper bound for $LS_{q}^{(k)}(I)$, and define the sketch sensitivity as follows.


Similar to the elastic sensitivity and residual sensitivity, our \textbf{Sketching Sensitivity (SKS)} must be smoothed using smooth sensitivity before it can be used with the Laplace mechanism.
\begin{equation}\label{equa:SKS}
  SKS=\max\limits_{k>0}e^{-\beta k}\hat{LS}_q^{(k)},
\end{equation}
where
\begin{equation}
  \hat{LS}_q^{(k)}=\max\limits_{I'\in \mathcal{I}^k}\max\limits_{i\in P}\frac{\mathop{mean}\limits_{s\in[1,s1]} \left|\prod_{j\in [n]-\{i\}}sk_s^{(k_j)}(R_j)\right|}{1-\tau}
\end{equation}

We use the following example to show how to compute the Sketching Sensitivity.
\begin{example}
  Consider the example database $I$ in Fig~\ref{Fig:intro_example}, we define two families of four-wise independent random variables for the join attributes $B$ and $C$ as $\xi_1$ and $\xi_2$.
We can construct sketches for $R_1$, $R_2$, and $R_3$ as follows.
$sk(R_1)= \sum_{t\in R_1} \xi_1(t[B])$,
$sk(R_2)=\sum_{t\in R_2} \xi_1(t[B])\cdot  \xi_2(t[C])$,
and $sk(R_3)=\sum_{t\in R_3} \xi_2(t[C])$.

The influence of adding a tuple to a database $I'$ with $k$ distance to the original database can be estimated by
\begin{equation}
\begin{split}
&\hat{LS}_q^{(k)}=\max\limits_{\sum_{j=1}^{3}|k_j|=k}\max\{
\mathop{mean}\limits_{s\in[1,s1]} \left|sk_s^{(k_1)}(R_1)\cdot sk_s^{(k_2)}(R_2)\right|,\\
&\mathop{mean}\limits_{s\in[1,s1]} \left|sk_s^{(k_1)}(R_1)\cdot sk_s^{(k_3)}(R_3)\right|,\\
&\mathop{mean}\limits_{s\in[1,s1]} \left|sk_s^{(k_2)}(R_2)\cdot sk_s^{(k_3)}(R_3)\right|\}\cdot\frac{1}{1-\tau}
\end{split}
\end{equation}
where $sk_s^{(k_j)}(R_j)=sk_s(R_j)+k_j$.
Then, the sketching sensitivity of $R_1\Join R_2\Join R_3$ can be computed according to Equation~\ref{equa:SKS}.
\end{example}


\section{Experiments}

We design experiments to verify the validity and efficiency of our methods. In section~\ref{sec:experimental-setup}, we introduce the experimental setup including the hardware, datasets, queries, competitors, error metrics and some parameters involved in the experiments. In section~\ref{sec:accuracy-experiments},  we compare the accuracy of our methods with RS and ES. And we also verify the efficiency of our methods to achieve differential privacy protection in section~\ref{sec:efficiency-experiments}. In section~\ref{sec:parameter-experiments}, we test the impact of different parameters such as privacy budget and sample rate. Finally, in section~\ref{sec:experiments-summary} we briefly summarize the experimental results.
\subsection{Experimental Setup}\label{sec:experimental-setup}
\begin{figure*}[htbp]
	\centering
	\subfigure[Q1]{
        \centering
		\begin{minipage}[b]{0.35\textwidth}
		\includegraphics[width=1\textwidth]{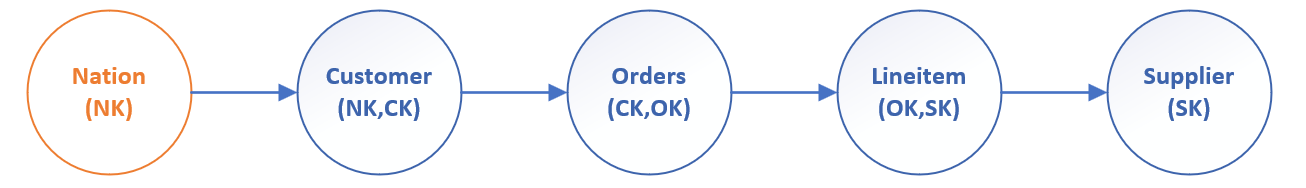}
		\end{minipage}
		\label{fig:Q1}
	}
    \subfigure[Q2]{
    	\centering
        \begin{minipage}[b]{0.26\textwidth}
   		\includegraphics[width=1\textwidth]{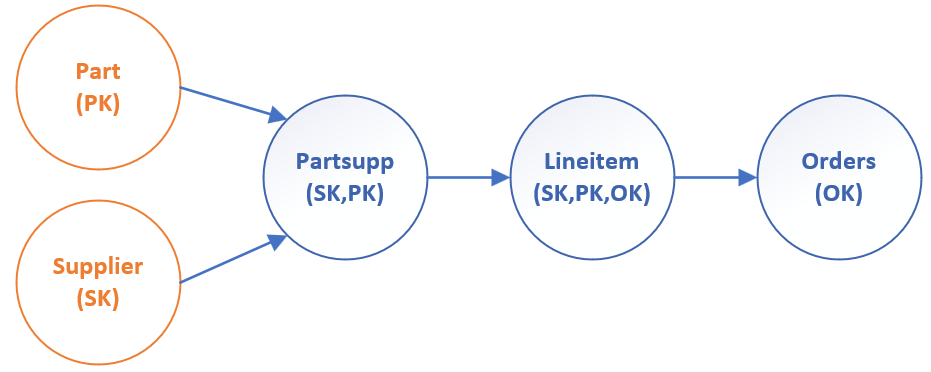}
    	\end{minipage}
	\label{fig:Q2}
    }
    \subfigure[Q3]{
    \centering
		\begin{minipage}[b]{0.26\textwidth}
		\includegraphics[width=1\textwidth]{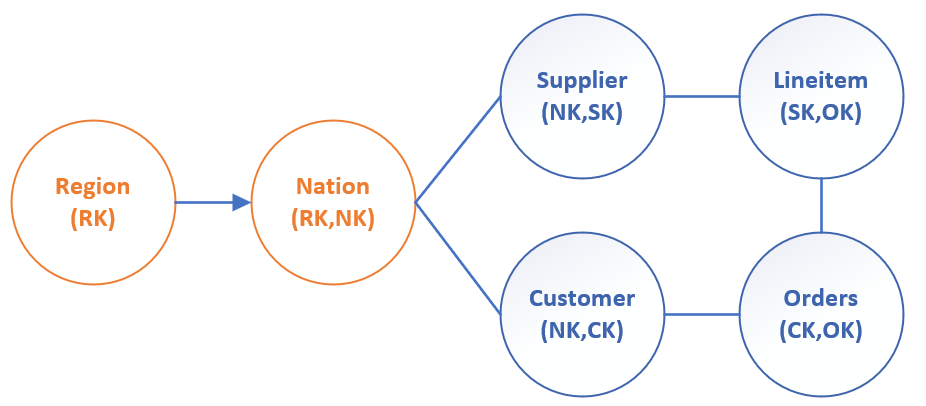}
		\end{minipage}
		\label{fig:Q1}
	}\\
    \subfigure[Q4]{
    \centering
    	\begin{minipage}[b]{0.3\textwidth}
   		\includegraphics[width=1\textwidth]{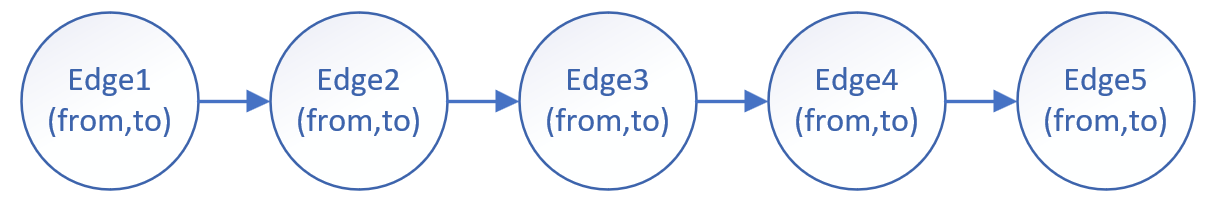}
    	\end{minipage}
	\label{fig:Q4}
    }
    \subfigure[Q5]{
    \centering
		\begin{minipage}[b]{0.17\textwidth}
		\includegraphics[width=0.7\textwidth]{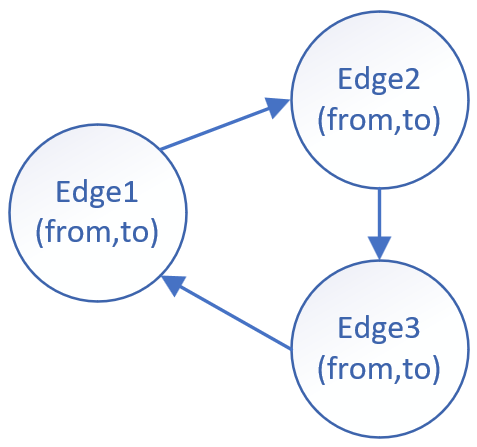}
		\end{minipage}
		\label{fig:Q5}
	}
    \subfigure[Q6]{
    \centering
    	\begin{minipage}[b]{0.17\textwidth}
   		\includegraphics[width=0.7\textwidth]{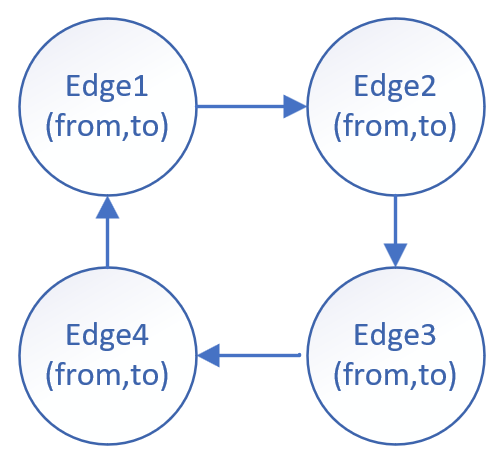}
    	\end{minipage}
	\label{fig:Q6}
    }
    \subfigure[Q7]{
    \centering
    	\begin{minipage}[b]{0.24\textwidth}
   		\includegraphics[width=0.7\textwidth]{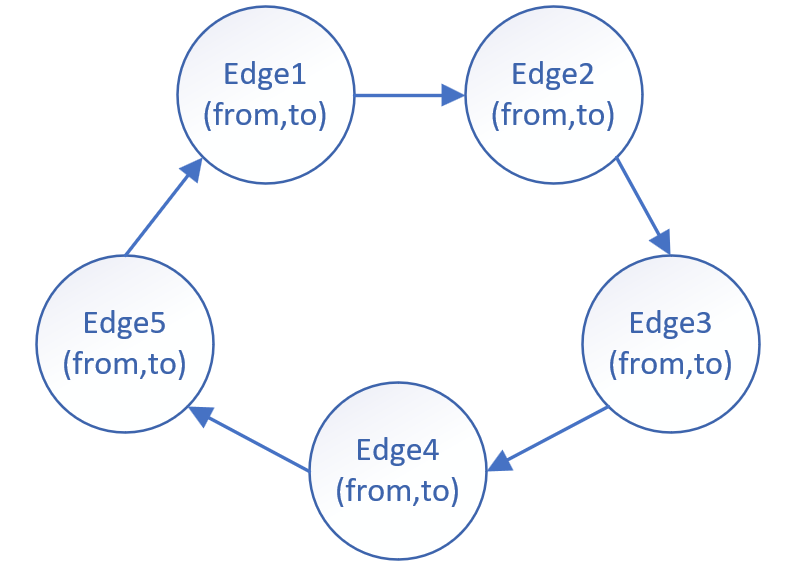}
    	\end{minipage}
	\label{fig:Q7}
    }
	\caption{Queries.}
	\label{fig:queries}
\end{figure*}
\noindent \underline{Hardware and Library}

All these experiments are implemented on a machine of 256 GB RAM running Ubuntu(20.04.1) with PostgreSQL(14.5) and Python 3.9.

\noindent \underline{Datasets}

We test our method based on two datasets, including TPC-H and Facebook ego-network.

\noindent\textbf{\textit{TPC-H dataset}}\footnote{https://www.tpc.org/tpch/}
The TPC-H dataset provides a set of schemas to support the TPC Benchmark H (TPC-H). It is used to measure the performance of highly-complex decision support databases. The dataset includes numerous schemas that only vary in the amount of data.The TPC-H schema consists of eight tables, each with a different number of columns and rows. The tables are named nation, region, part, customer, lineitem, orders, partsupp, and supplier. The schema is designed to be representative of real-world decision support systems. Due to the included data and scale of each table, we treat the first three tables as public relations and the other five tables as private relations.Besides, for convenience, we extract the key attributes for join condition from the dataset to create 8 tables correspondingly.

\noindent\textbf{\textit{Facebook ego-network dataset.}}\footnote{https://snap.stanford.edu/data/ego-Facebook.html}
For the reason that the TPC-H datasets is evenly distributed, so we introduce another dataset with skewness.Facebook ego-network dataset contains 4,039 nodes and 176,467 edges. As is shown in ~\cite{Dong2021ResidualSF}, we use similar ways to merge this dataset into 5 relations, which all consists two attributes, e.g. from and to.

\noindent \underline{Queries}

The queries for the experiments are shown in Figure~\ref{fig:queries}. We use Q1-Q3 for TPC-H dataset, and Q4-Q7 for Facebook dataset. The queries include chain queries, acyclic queries and cyclic queries. The orange circles represent non-private relations, and the blue circles represent the private relations.

\noindent \underline{Competitors}

In the following experiments, our methods are called ``Sampling-SE'' and ``Sketch-SE'', which stand for sampling-based sensitivity estimation and sketch-based sensitivity estimation, respectively.
The competitors of our methods include ES and RS.

(1) ES: Elastic Sensitivity is a smooth upper bound of local sensitivity. ES is computed based on the frequency of the most frequent join attribute in each relation.

(2) RS: Residual Sensitivity is a smooth upper bound tighter than ES. It is computed based on the maximum boundaries of the residual queries.

\noindent \underline{Error Metrics}

Deviation(DE):
\par $DE = |TrueResult-EstimateResult|$.

\noindent \underline{Parameters}
\par $\epsilon$: the privacy budget, symbolizes the level of privacy protection.
\par $\delta$: a new privacy parameter in ($\epsilon ,\delta$)-DP representing the failure probability of meeting pure differential privacy.
\par $r$: the sampling rate.

\subsection{Accuracy}\label{sec:accuracy-experiments}
\begin{figure}[htbp]
	\centering
	\subfigure[Q1]{
        \centering
		\begin{minipage}[b]{0.3\textwidth}
		\includegraphics[width=1\textwidth]{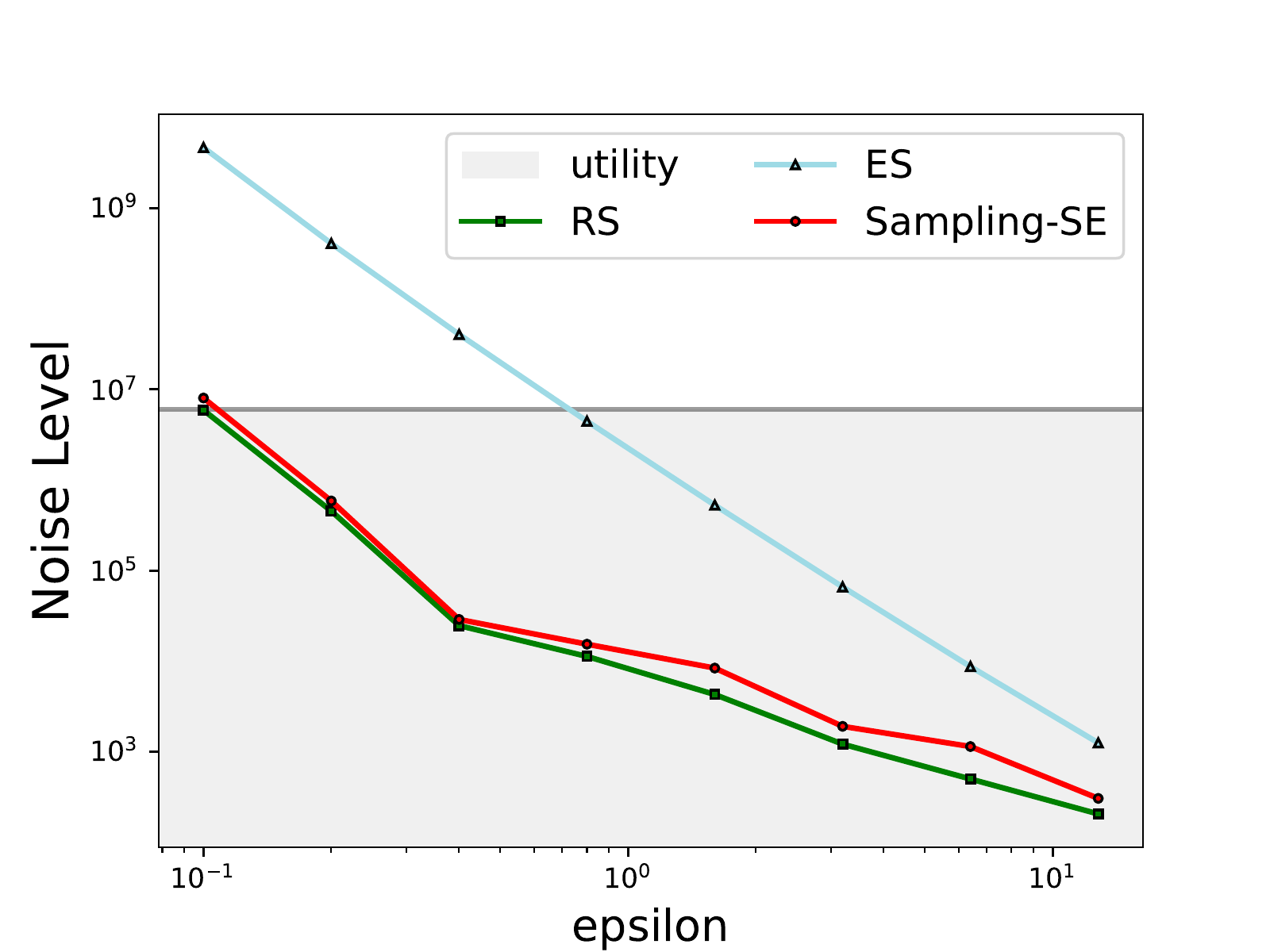}
		\end{minipage}
		\label{fig:epsilon_q1}
	}
   \subfigure[Q2]{
    \centering
    	\begin{minipage}[b]{0.3\textwidth}
   		\includegraphics[width=1\textwidth]{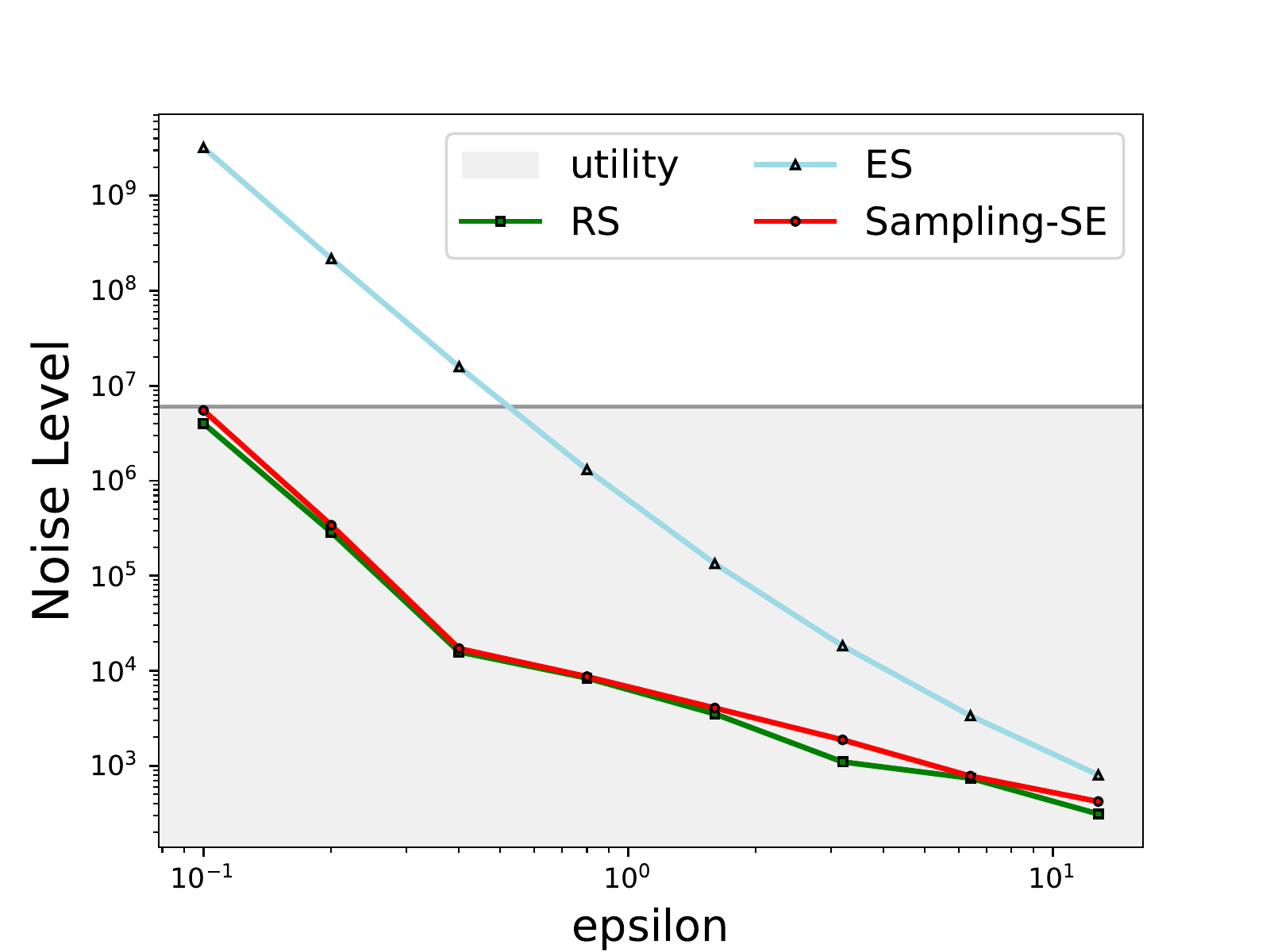}
    	\end{minipage}
	\label{fig:epsilon_q2}
    }
    \subfigure[Q3]{
    \centering
    	\begin{minipage}[b]{0.3\textwidth}
   		\includegraphics[width=1\textwidth]{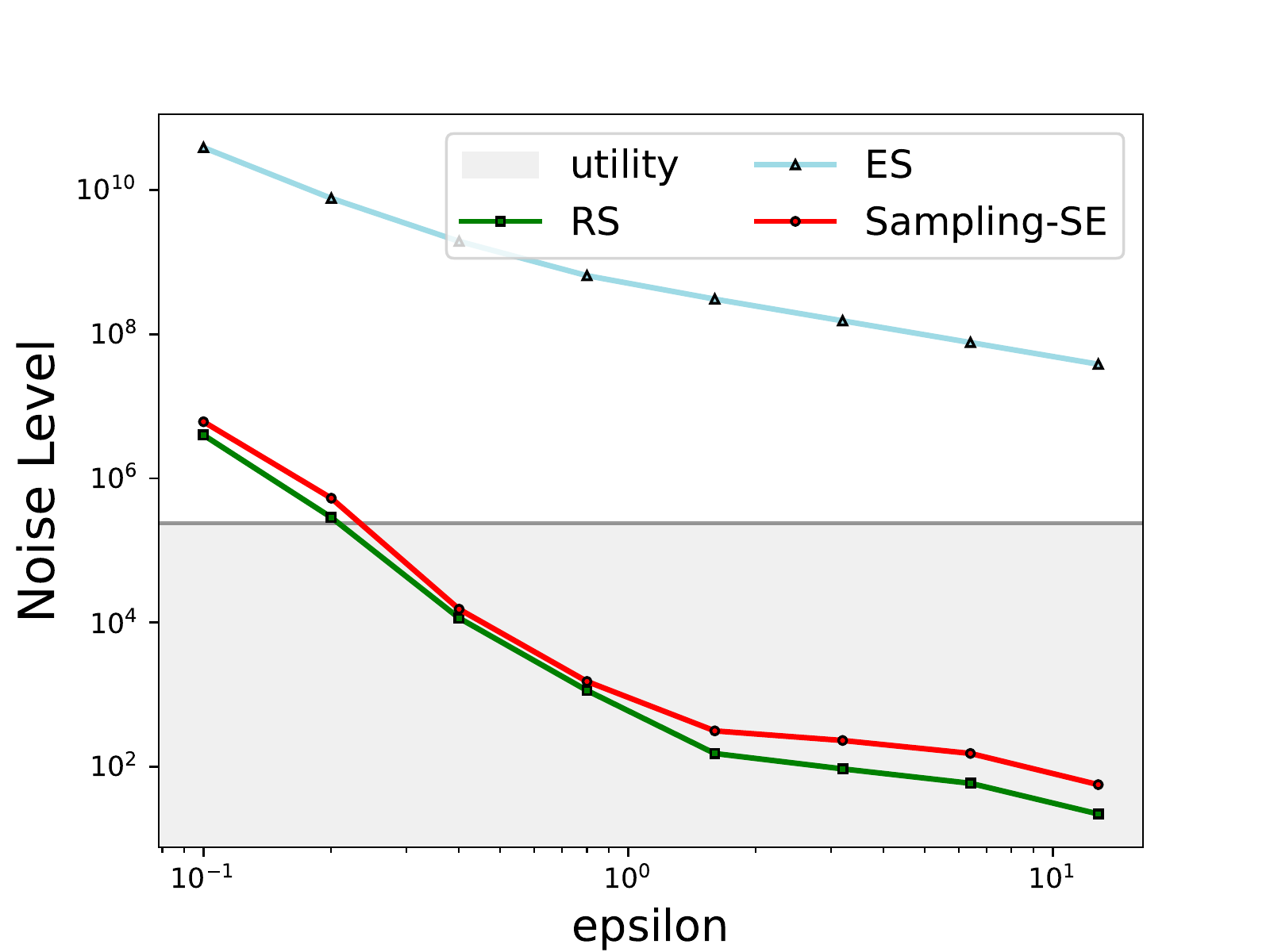}
    	\end{minipage}
	\label{fig:epsilon_q3}
    }
	\caption{Impact of $\epsilon$ on TPC-H dataset.}
	\label{fig:epsilon_TPCH}
\end{figure}

\begin{figure}[htbp]
	\centering
	\subfigure[Q4]{
        \centering
		\begin{minipage}[b]{0.38\textwidth}
		\includegraphics[width=1\textwidth]{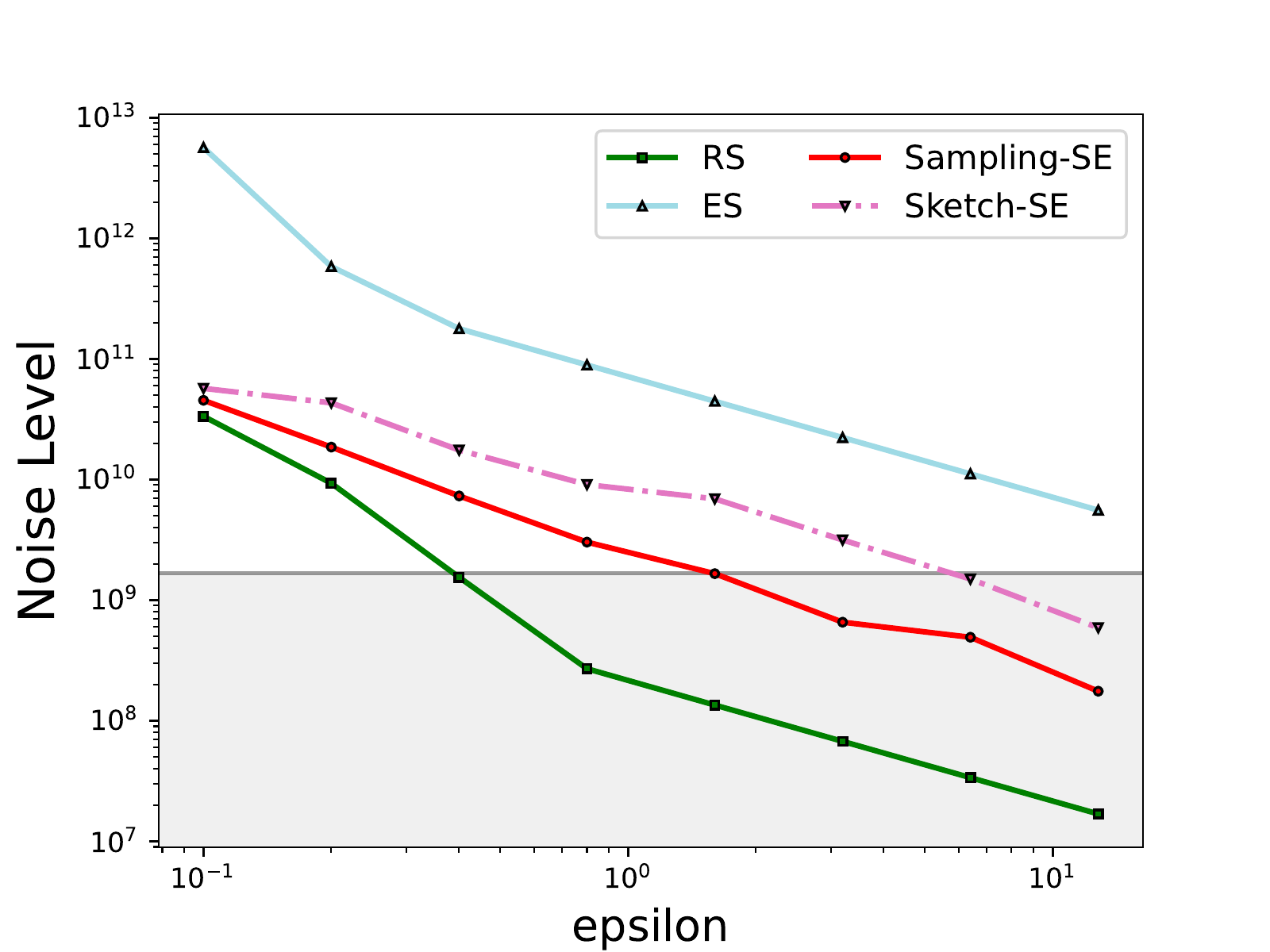}
		\end{minipage}
		\label{fig:epsilon_q4}
	}
   \subfigure[Q5]{
    \centering
    	\begin{minipage}[b]{0.38\textwidth}
   		\includegraphics[width=1\textwidth]{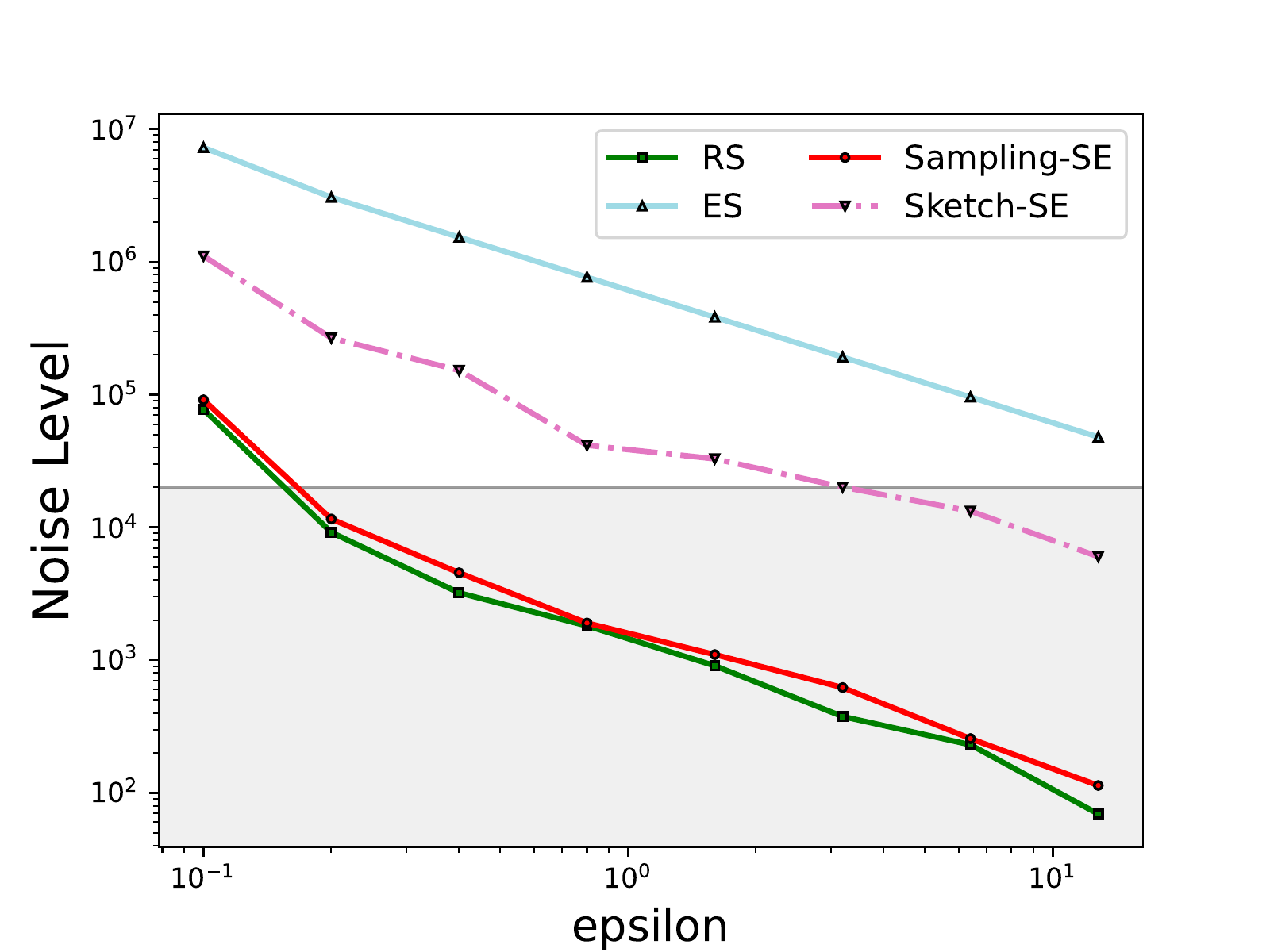}
    	\end{minipage}
	\label{fig:epsilon_q5}
    }

    \subfigure[Q6]{
    \centering
    	\begin{minipage}[b]{0.38\textwidth}
   		\includegraphics[width=1\textwidth]{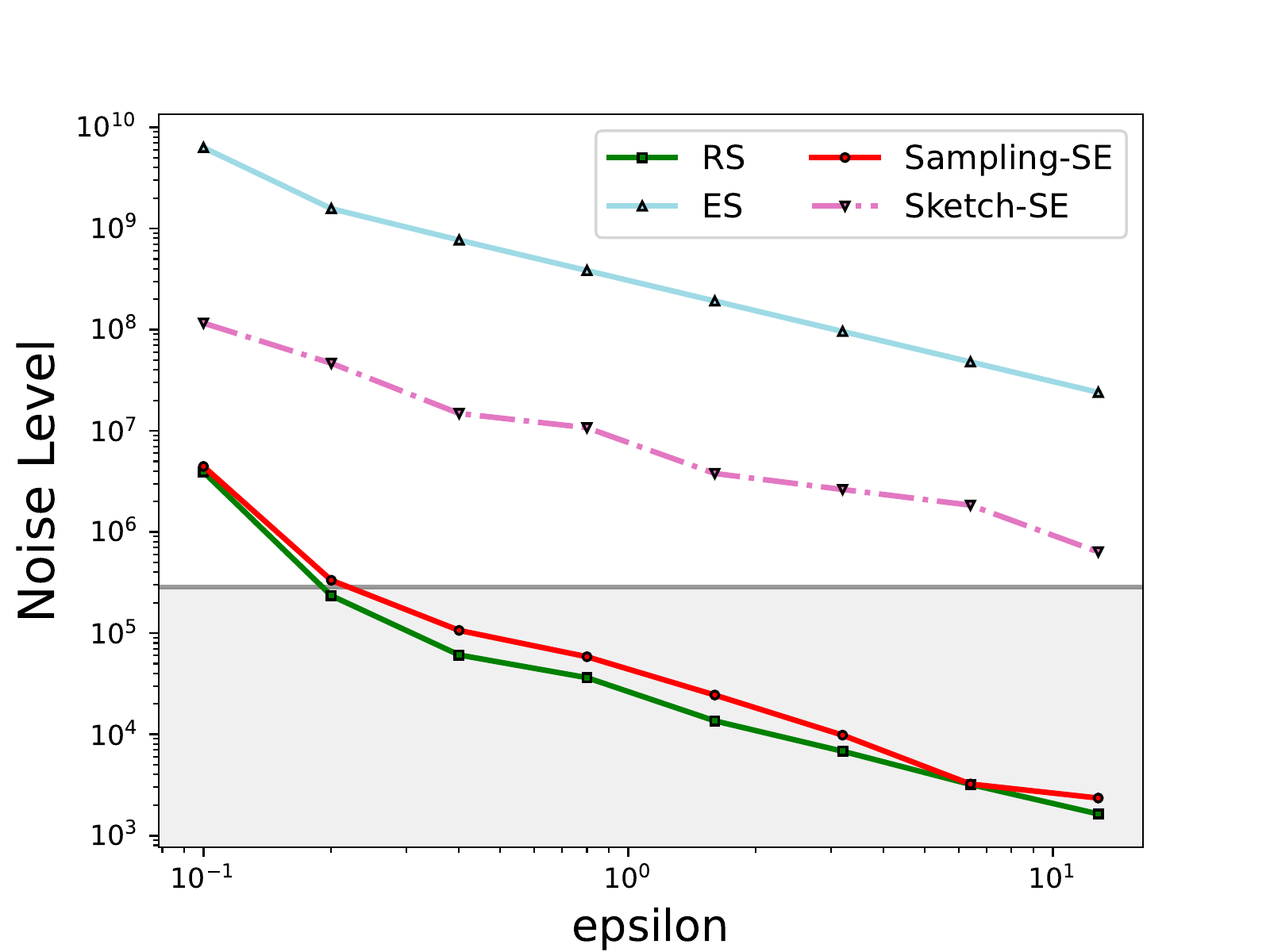}
    	\end{minipage}
	\label{fig:epsilon_q6}
    }
    \subfigure[Q7]{
    \centering
    	\begin{minipage}[b]{0.38\textwidth}
   		\includegraphics[width=1\textwidth]{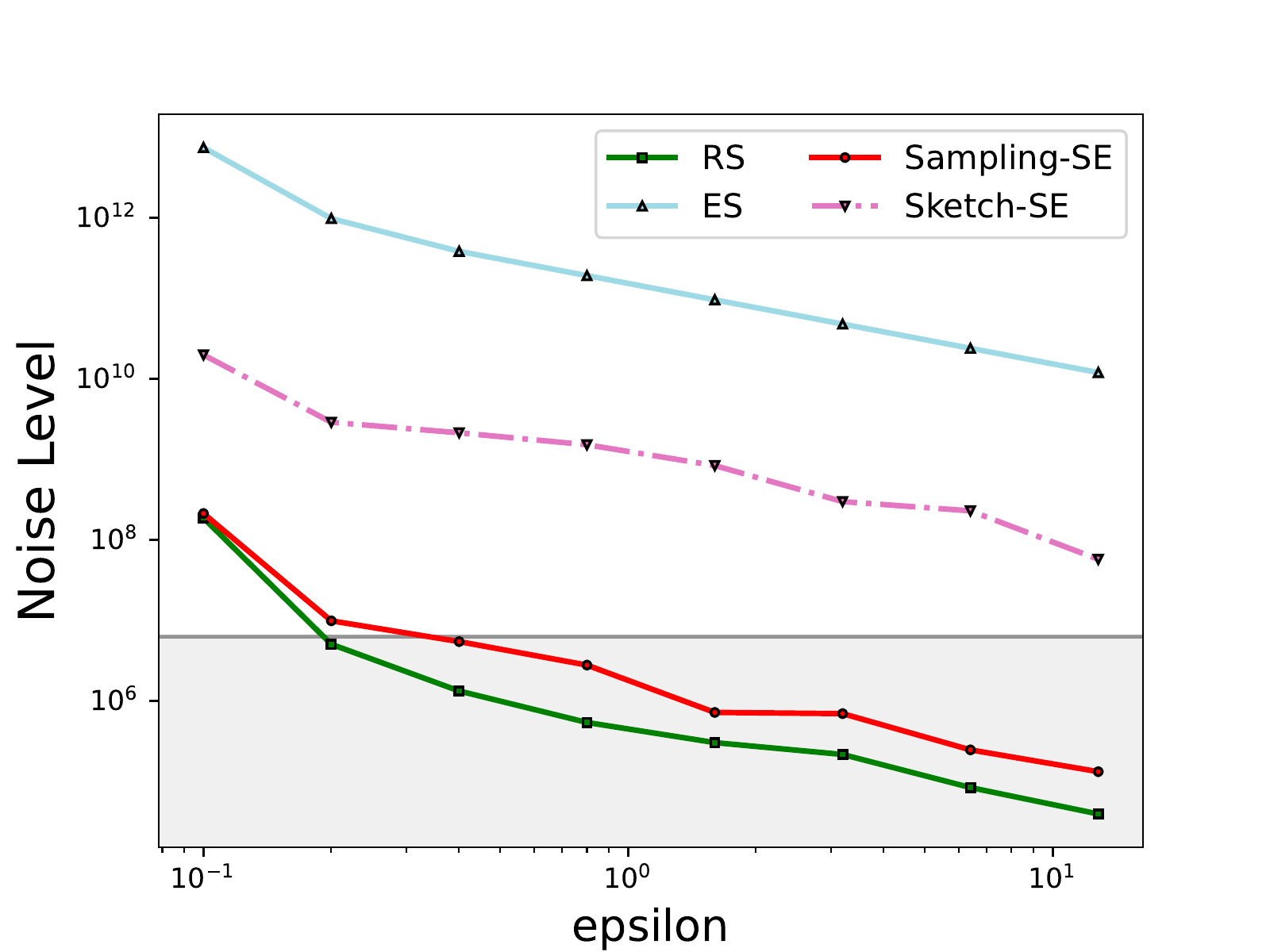}
    	\end{minipage}
	\label{fig:epsilon_q7}
    }
	\caption{Impact of $\epsilon$ on Facebook dataset.}
	\label{fig:epsilon_Facebook}
\end{figure}

In the following experiments, our methods are called ``Sampling-SE'' and ``Sketch-SE'', which stand for sampling-based sensitivity estimation and sketch-based sensitivity estimation, respectively. We compare the accuracy of our Sampling-SE and Sketch-SE with ES and RS on the Facebook dataset. And on TPC-H dataset, we only compare the Sampling-SE with ES and RS. Because the result of TPC-H involves too much distinct values, and the sketch-based method is not suitable for the join result with a large domain. We tested the noise level of different methods with $\epsilon$ in [0.1, 0.2, 0.4, 0.8, 1.6, 3.2, 6.4, 12.8]. Smaller $\epsilon$ leads to higher privacy protection and more noise. We set the sampling rate as 0.0001 for Sampling-SE, and set number of estimators for each relation as 100,000 for Sketch-SE.

The experimental results on TPC-H(1GB) and Facebook are shown in the fig~\ref{fig:epsilon_TPCH} and fig~\ref{fig:epsilon_Facebook}, respectively. The error is measured by the deviation from the true query result, i.e., the noise added. In these figure we also plot the actual query result, i.e., the shaded area in the figure. It is easy to see that the value of noise level below query result is considered to have utility, and the above value is considered to be out of utility correspondingly.

We can learn from the figures that our method Sampling-SE has low noise level similar to RS, and they are both much smaller than the noise level computed according to ES, which further highlights the advantages of our sampling-based estimation strategy. Figure~\ref{fig:epsilon_Facebook} illustrates the noise level on Facebook dataset. The noise level of Sketch-SE is higher than Sampling-SE but lower than ES. As the domain of the join result increases with the number of the relations involved in the query, the number of estimators required to provide a sufficiently accurate result is quite large. Thus, Sketch-SE is less accurate than the Sampling-SE when the join query involve many relations with a large number of distinct join values. However, Sketch-SE is still more accurate than ES as shown in fig~\ref{fig:epsilon_Facebook}.

\subsection{Efficiency}\label{sec:efficiency-experiments}

In this part we verify the efficiency of our method compared to RS and ES. We measure the efficiency in terms of the time between submitting the query and returning the noised query results, which consists of two main components: the time of answering the query, and the time of computing the sensitivity and adding noise according to the differential privacy mechanism. The time of computing the maximum frequency of each join attribute for ES and the time of constructing sketches for each relation is not included in the figures, since they are computed offline.



\begin{figure*}[htbp]
	\centering
	\subfigure[Q1]{
        \centering
		\begin{minipage}[b]{0.3\textwidth}
		\includegraphics[width=1.2\textwidth]{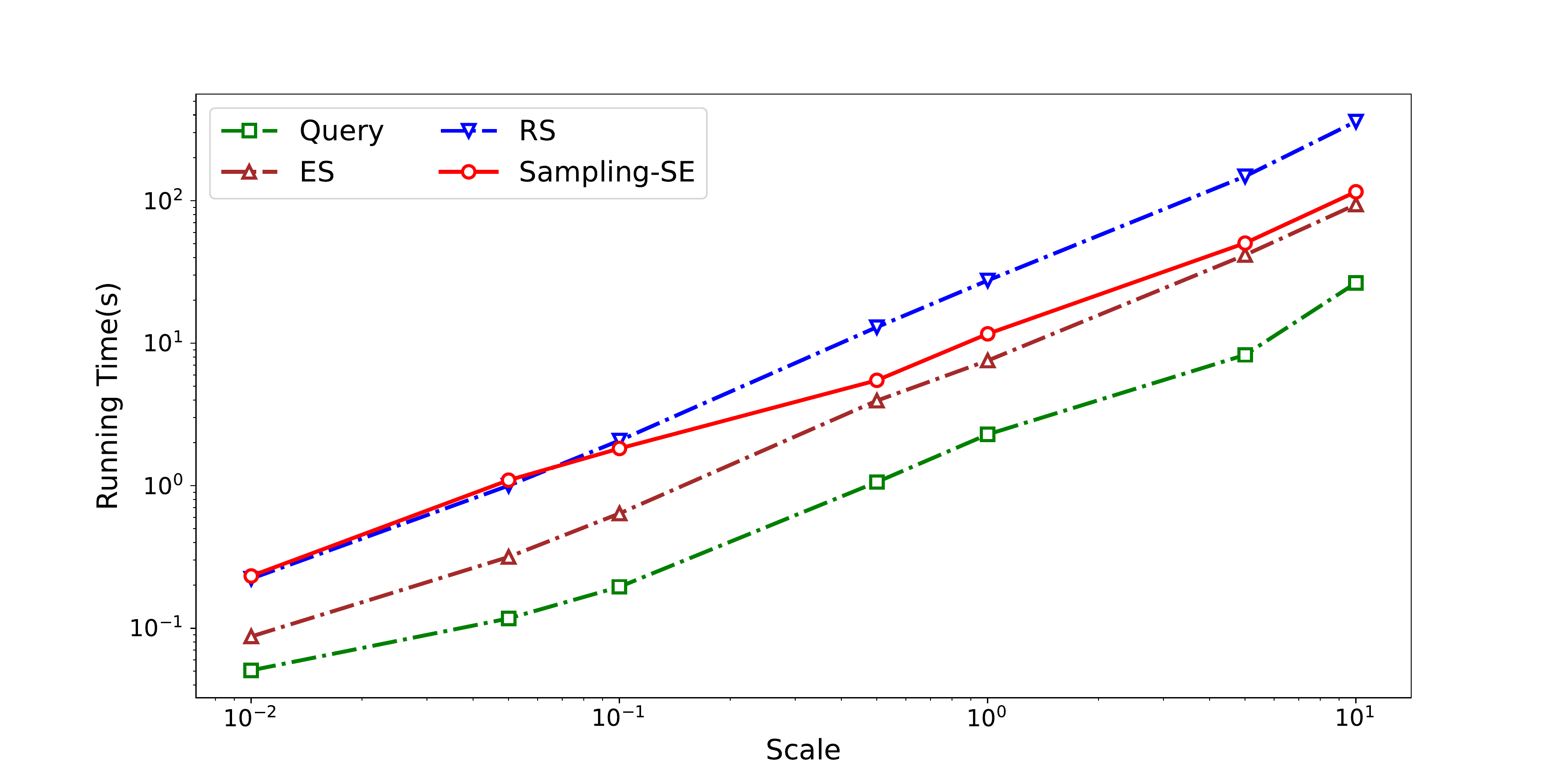}
		\end{minipage}
		\label{fig:scale_time_q1}
	}
   \subfigure[Q2]{
    \centering
    	\begin{minipage}[b]{0.3\textwidth}
   		\includegraphics[width=1.2\textwidth]{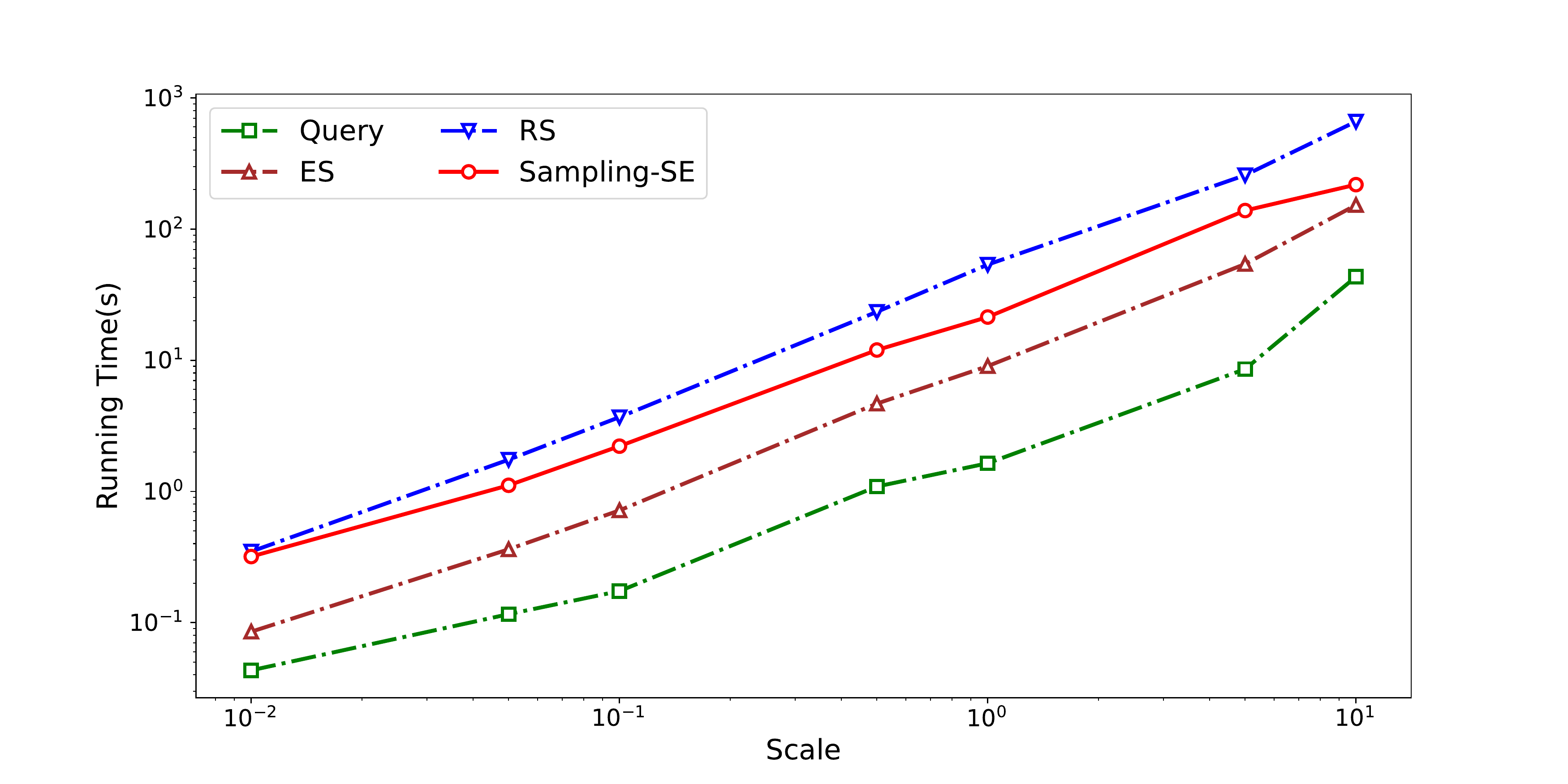}
    	\end{minipage}
	\label{fig:scale_time_q2}
    }
    \subfigure[Q3]{
    \centering
    	\begin{minipage}[b]{0.3\textwidth}
   		\includegraphics[width=1.2\textwidth]{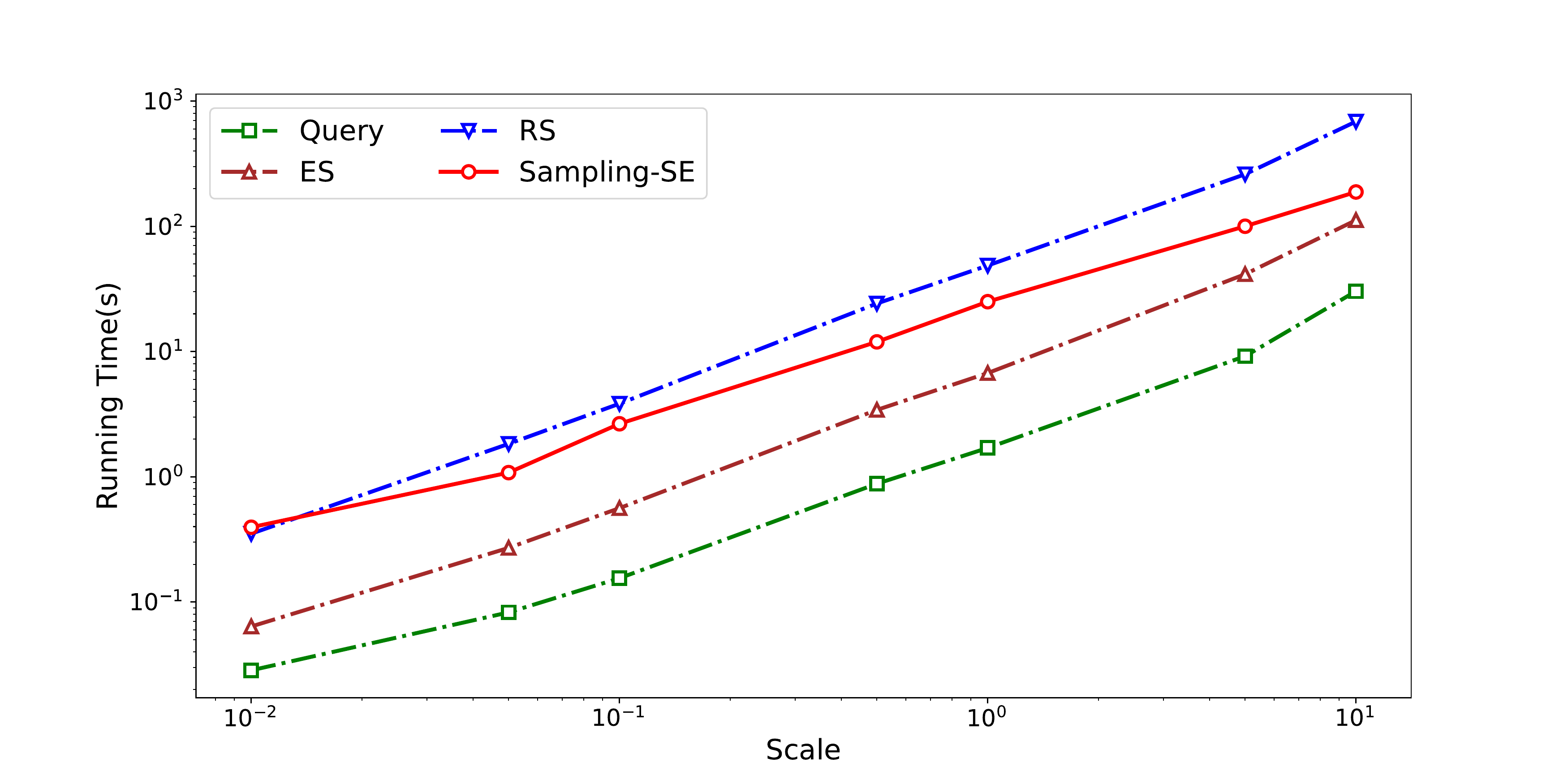}
    	\end{minipage}
	\label{fig:scale_time_q3}
    }
	\caption{Running time of Queries on TPC-H.}
	\label{fig:scale_time_TPCH}
\end{figure*}

Figure~\ref{fig:scale_time_TPCH} shows the running times of RS and ES for different queries and data scales on TPC-H dataset, where $\epsilon$=0.8 for all three queries and $\delta =10^{-7}, 2\times 10^{-8}, 10^{-8}, 2\times 10^{-9}, 10^{-9}, 2\times 10^{-10}, 10^{-10}$ for different scales including 0.01G, 0.05G, 0.1G, 0.5G, 1G, 5G, 10G respectively on each query. Apparently, the time overhead of Sampling-SE is always somewhere between RS and ES. The effect is particularly noticeable at larger data sizes.
This discrepancy stems mainly from two reasons. On the one hand, Sampling-SE reduces the cost of computing many residual queries by sampling, on the other hand, Some complex residual queries with primary-foreign joins can be reduced to find the maximum frequency value of one relation, which further reduced the cost.

\begin{figure}[htbp]
  \centering
  \includegraphics[width=24em]{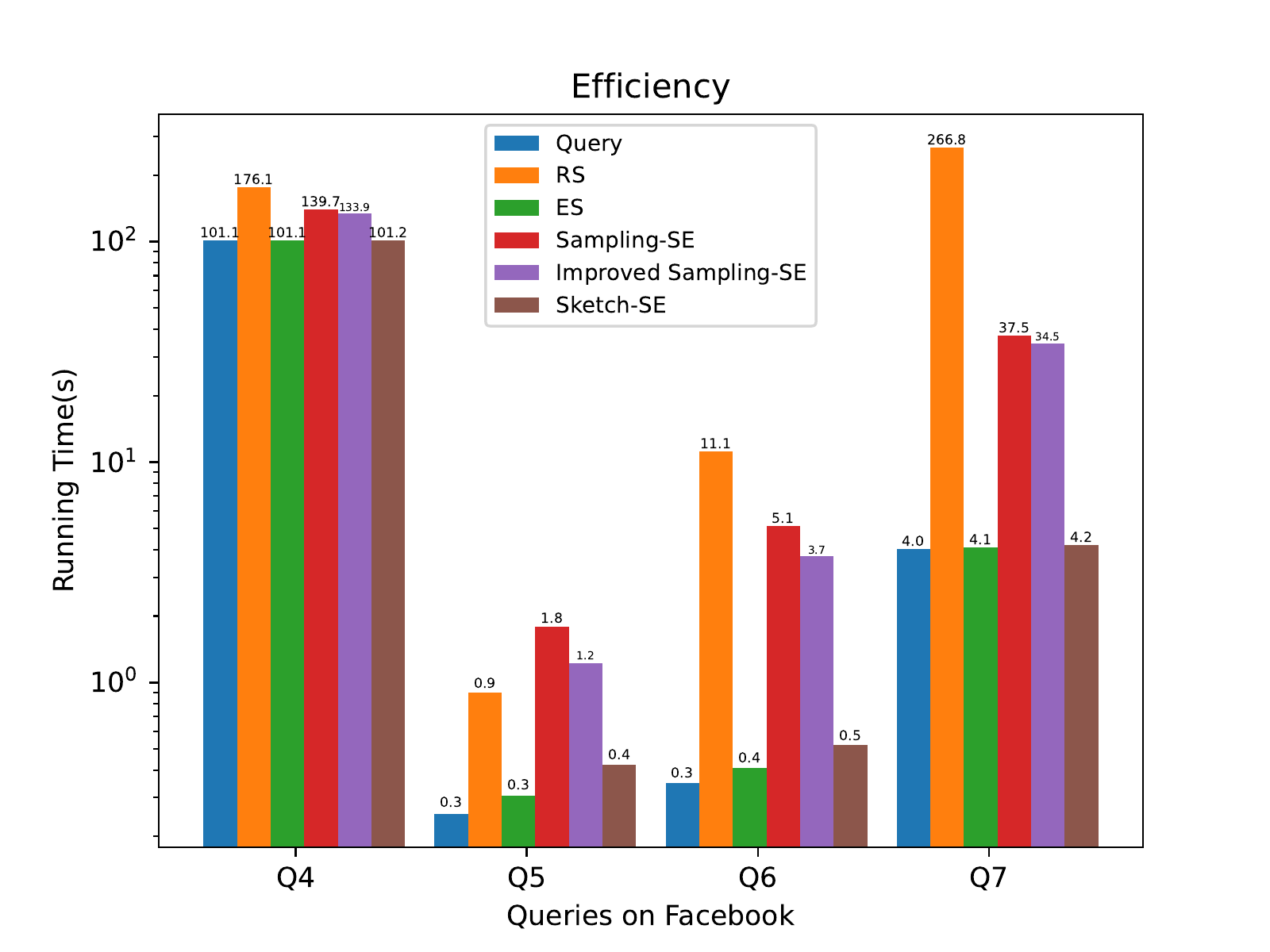}
  \caption{Running time of Queries on Facebook.}
  \label{figure:scale_time_Facebook}
\end{figure}

We also compared the efficiency of different method on Q4 to Q7, and the results are plotted in Fig~\ref{figure:scale_time_Facebook}. We set the parameters $\epsilon=0.8$ and $\delta=10^{-7}$. For our Sampling-SE method we fix the sample rate at 0.0001.

It can be seen that the Sampling-SE also works well on the Facebook dataset for most cases. The Sampling-SE is slightly less efficient than RS on Q5 due to the fact that only three small relations are included in this query, and the time to compute RS of Q5's is inherently short. But our Sampling-SE approach works much better with larger data sizes and complex join conditions where the utility gains are more pronounced. Since the statistics for ES and the sketches for Sketch-SE are computed offline, the online efficiency of them are much more efficient than other methods.

\subsection{Impact of parameters}\label{sec:parameter-experiments}
\underline{Impact of Sample Size}
\begin{figure}[htbp]
  \centering
  \includegraphics[width=0.5\textwidth]{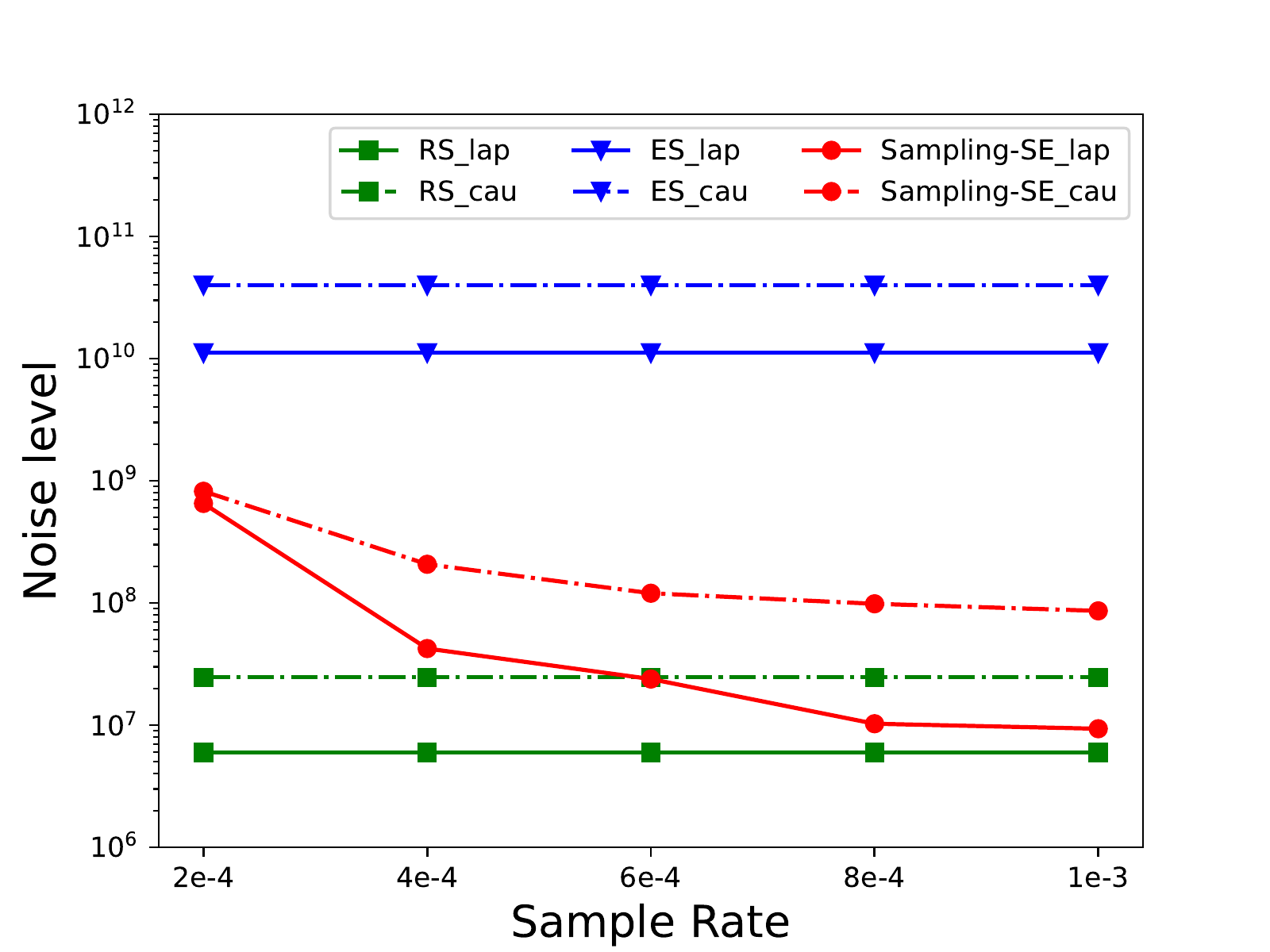}
  \caption{Noise level under different sample rate.}
  \label{fig:SampleRate_noise}
\end{figure}
The key factor of our Sampling-SE method is the sample rate. We conducted experiments to see how the sample rate $r$ affects the noise results. For simplicity, we take two datasets and two chain queries Q1 and Q4 in this experiment. For the parameters, we set scale=1 GB, $\epsilon$ =6.4, and $\delta =2\times 10^{-10}$ for TPC-H while $\delta =10^{-7}$ for Facebook. Besides, different noise mechanisms including $Laplace Mechanism$ ~\cite{Dwork2006CalibratingNT} and $General Cauchy Mechanism$ ~\cite{Nissim2007SmoothSA} are implemented to confirm our results. Significant results are illustrated in Fig~\ref{fig:SampleRate_noise} and Fig~\ref{fig:SampleRate_result}.

\begin{figure}[htbp]
  \centering
  \includegraphics[width=0.5\textwidth]{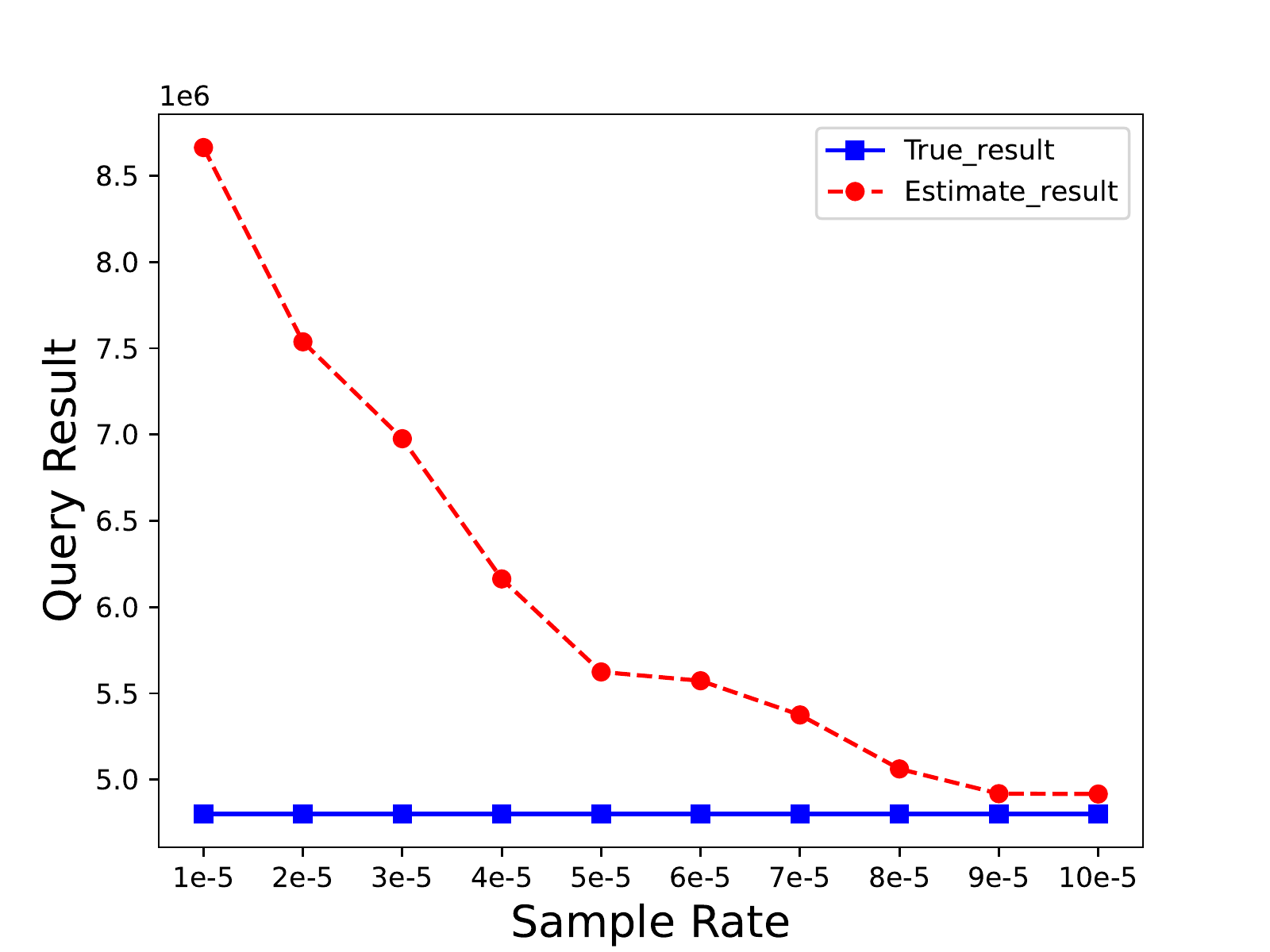}
  \caption{Estimate result under different sample rate.}
  \label{fig:SampleRate_result}
\end{figure}

As shown in Fig~\ref{fig:SampleRate_noise}, the Sampling-SE with different differential privacy mechanism is much more accurate than ES. 
The noise level of Sampling-ES becomes much closer to the accurate result, as the sampling size increases. Moreover, the results under the Laplace mechanism are consistently smaller than those under the GeneralCauchy mechanism for both the two baselines and our method.

To further validate the effectiveness of the sampling method, we consider a residual query: "select max(count) from (select edge2\_from, count(*) from edge2, edge3, edge4, edge5 where edge2\_to=edge3\_from and edge3\_to=edge4\_from and edge4\_to=edge5\_from group by edge2\_from) as t;", which is also a typical query structure that our Sampling-SE aiming at. Fig~\ref{fig:SampleRate_result} shows the change in query estimation results as the sampling rate varies from $10^{-5}$ to $10^{-4}$. It can be observed that with a relatively small sampling rate we can obtain a result with high accuracy.

\begin{figure}[htbp]
  \centering
  \includegraphics[width=0.5\textwidth]{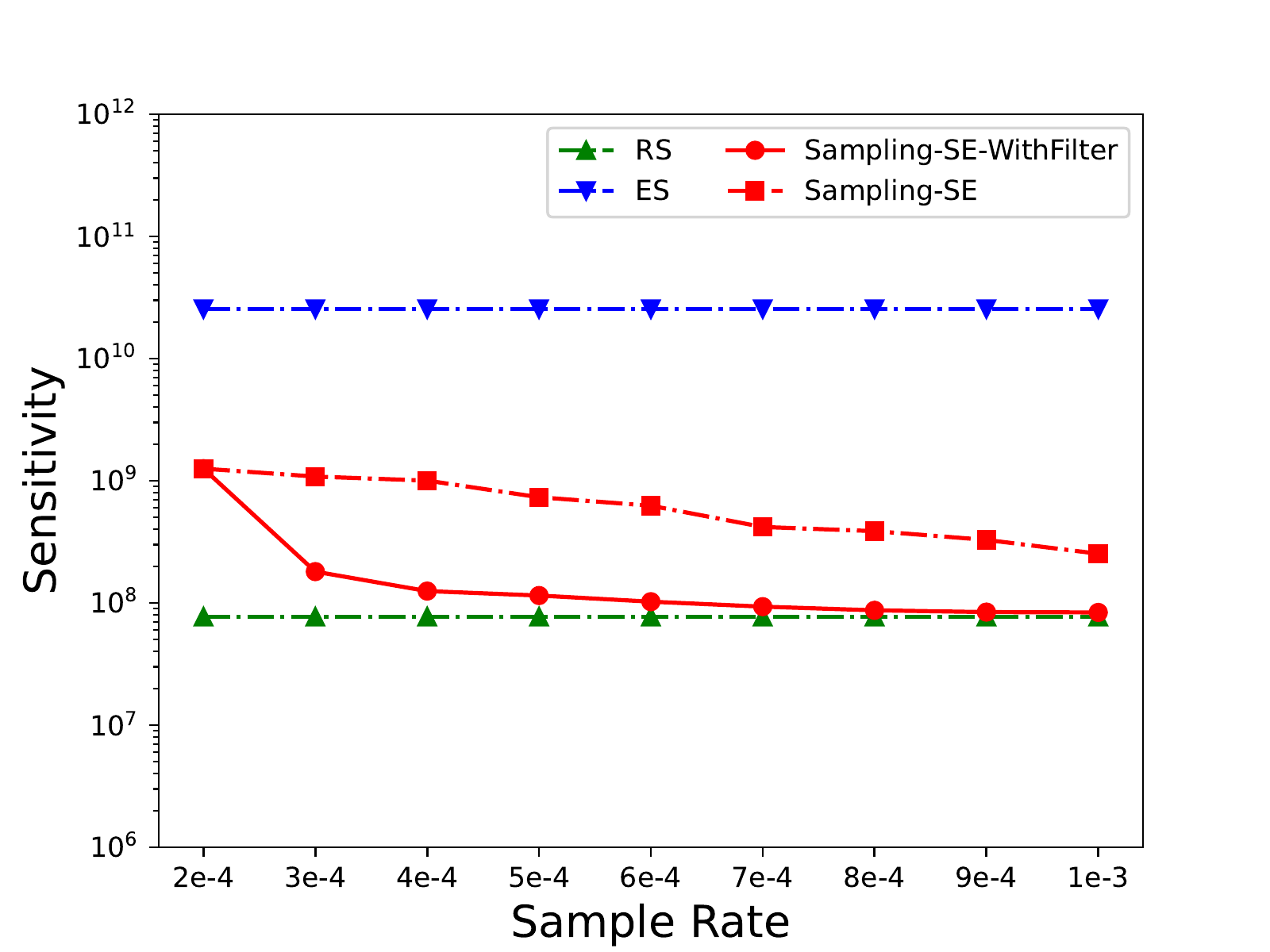}
  \caption{Sensitivity under different sample rate.}
  \label{fig:SampleRate_sensitivity}
\end{figure}

Due to the fact that the noise of differential privacy is associated to the privacy budget $\epsilon$, and a small value of it causes a large noise variation. Therefore, we tested the effect of sampling rate on sensitivity as a way to eliminate the interference of the differential privacy noise added on the data.  We use the same conditions as shown in the experiment on Q4 earlier. Here we also set the privacy budget $\epsilon$ =6.4 and $\delta =10^{-7}$ on Q4 using Laplace mechanism. The experimental result is shown in Fig~\ref{fig:SampleRate_sensitivity}.
To deal with the join results including too much distinct values and further improve the efficiency of our Sampling-SE algorithm, we use a strategy of estimating the largest group size among the groups included in some random samples of the query result instead of estimating the size of each group. We call this strategy Sampling-SE-WithFilter as shown in Fig~\ref{fig:SampleRate_sensitivity}.
We can observe that Sampling-SE-WithFilter converges faster than Sampling-SE as the sampling rate increases.


\underline{Impact of Data scale}

\begin{figure*}[htbp]
	\centering
	\subfigure[Q1]{
        \centering
		\begin{minipage}[b]{0.3\textwidth}
		\includegraphics[width=1\textwidth]{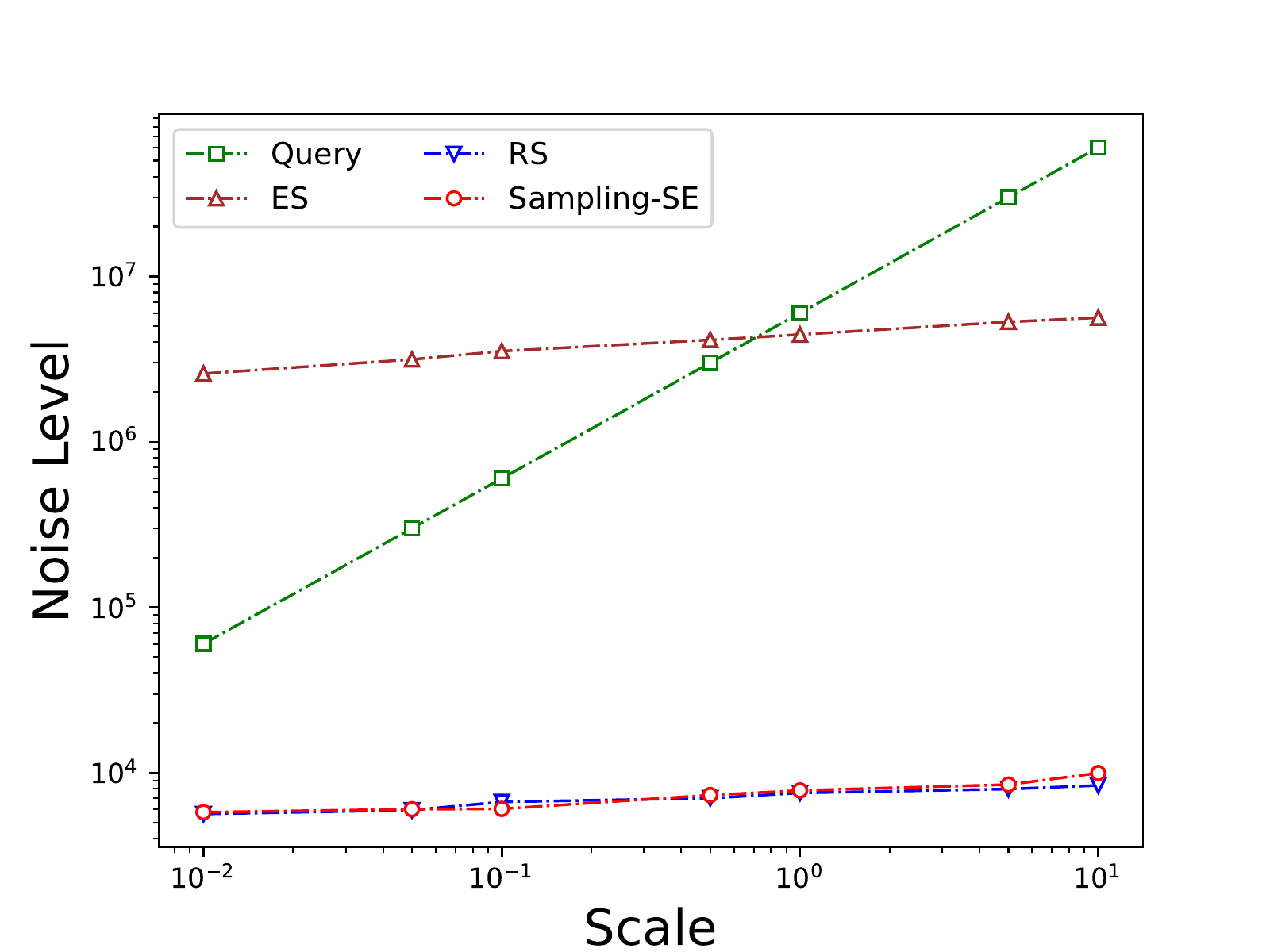}
		\end{minipage}
		\label{fig:Scale_noise_q1}
	}
   \subfigure[Q2]{
    \centering
    	\begin{minipage}[b]{0.3\textwidth}
   		\includegraphics[width=1\textwidth]{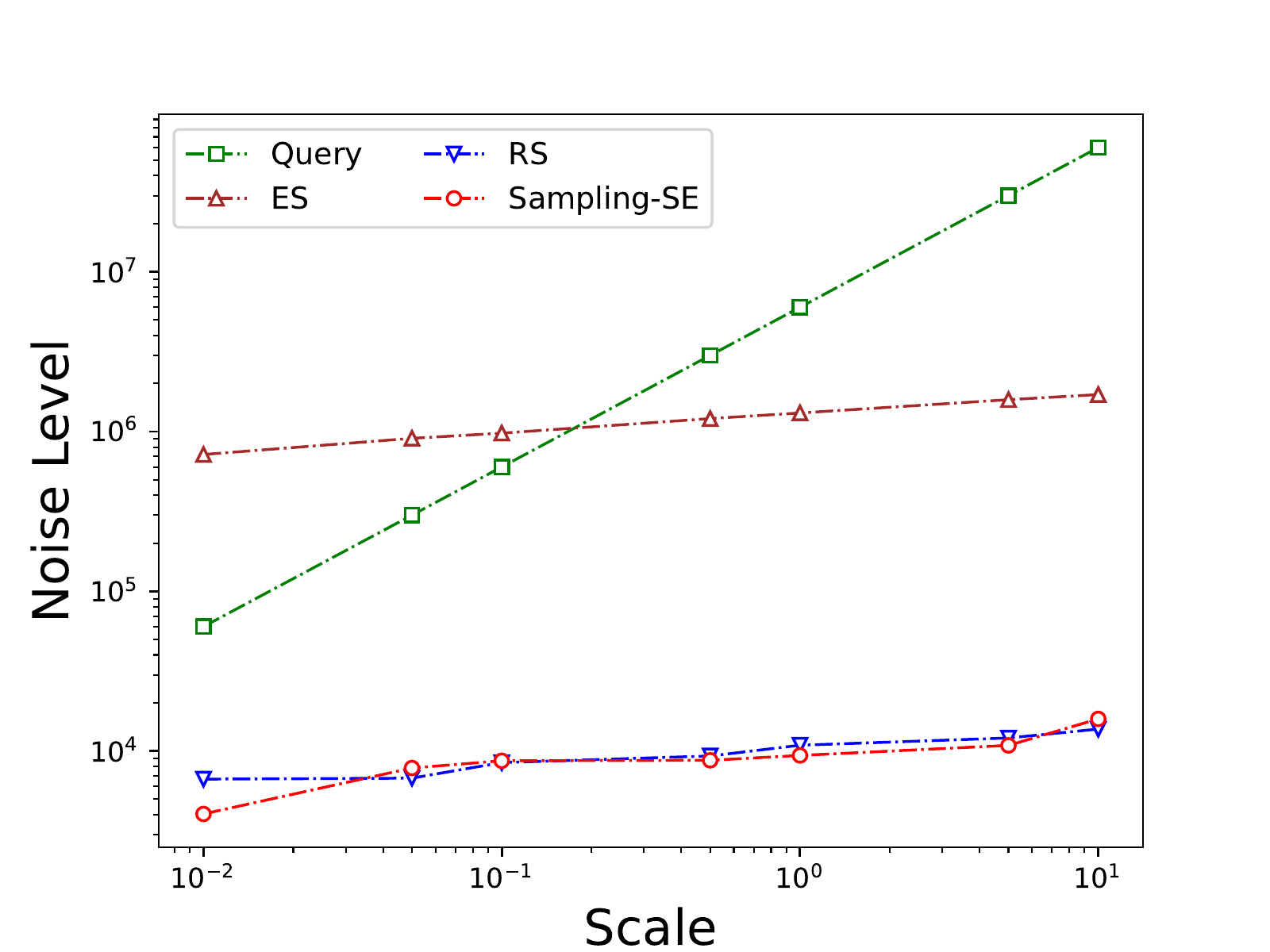}
    	\end{minipage}
	\label{fig:Scale_noise_q2}
    }
    \subfigure[Q3]{
    \centering
    	\begin{minipage}[b]{0.3\textwidth}
   		\includegraphics[width=1\textwidth]{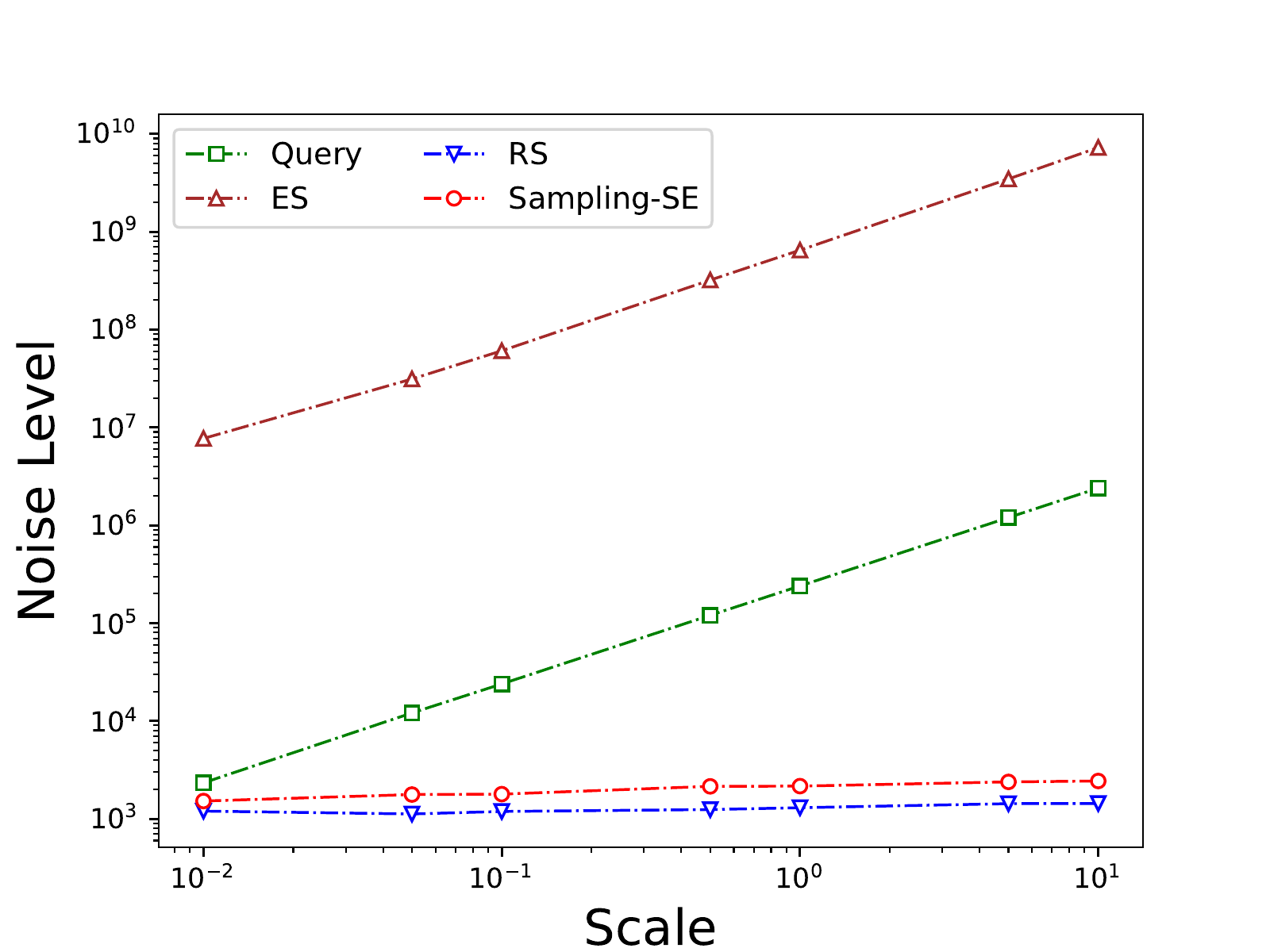}
    	\end{minipage}
	\label{fig:Scale_noise_q3}
    }
	\caption{Impact of data scale on the noise level.}
	\label{fig:Scale_noise_TPCH}
\end{figure*}


In this part we evaluated the effects of data scale of TPC-H on the noise level. We use TPC-H datasets with scale factors ranging from 0.01 to 10 and test the performance of different methods with Q1,Q2,and Q3. For our method we fix the sampling rate at 0.0001. We can learn from Fig~\ref{fig:Scale_noise_TPCH} that both our Sample-SE and RS produce less fluctuation with increasing scale, meaning that our method is suitable for large scale datasets. It is notable that the maximum boundaries of Q1 and Q2 do not change after using Sampling-SE method, so the noise level of these two queries is almost identical to the original RS result.
We compared these noise levels to the query answer result to verify the utility with noise added which means that a noise level higher than the original query result can completely obscure the information of it. Particularly, the noise levels from RS and Sampling-SE are always lower than query answer.

\subsection{Summary for experimental results}\label{sec:experiments-summary}

The experimental results are summarized as follows:
\begin{itemize}
  \item [$\bullet$] Sampling-SE and Sketch-SE both have higher efficiency than Residual Sensitivity for join queries with large scale datasets.

  \item [$\bullet$] Sampling-SE with an appropriate sample rate has the equal level of accuracy to Residual Sensitivity while keeping a lower time overhead.

  \item [$\bullet$] Sketch-SE has the same level of efficiency as Elastic Sensitivity but results in a relatively lower value of sensitivity which leads to higher accuracy.
\end{itemize}

\section{Conclusion}
In this paper, we proposed two sensitivity estimation methods for multi-join queries based on sampling and sketches, respectively. The sampling-based method gives an estimate of sensitivity similar to the residual sensitivity with higher efficiency, and the sketch-based method makes  a more accurate sensitivity but similarly efficient sensitivity estimation than the elastic sensitivity. We will adopt the idea of sensitivity estimation methods for more kinds of complex queries including complex predicates and user-defined functions in the future.

\section*{Acknowledgements}

This work was supported by NSFC grant 62202113, 
SDGC2229, SL2022A04J01306,
the Major Key Project of PCL (Grant No.PCL2021A09, PCL2021A02, PCL2022A03).



\bibliographystyle{elsarticle-num}
\bibliography{mybibfile}

\begin{thebibliography}{10}
\expandafter\ifx\csname url\endcsname\relax
  \def\url#1{\texttt{#1}}\fi
\expandafter\ifx\csname urlprefix\endcsname\relax\def\urlprefix{URL }\fi
\expandafter\ifx\csname href\endcsname\relax
  \def\href#1#2{#2} \def\path#1{#1}\fi

\bibitem{Johnson2017TowardsPD}
N.~M. Johnson, J.~P. Near, D.~X. Song, Towards practical differential privacy
  for sql queries, Proc. VLDB Endow. 11 (2017) 526--539.

\bibitem{Dong2021ResidualSF}
W.~Dong, K.~Yi, Residual sensitivity for differentially private multi-way
  joins, Proceedings of the 2021 International Conference on Management of Data
  (2021).

\bibitem{Dwork2006DifferentialP}
C.~Dwork, Differential privacy, in: Encyclopedia of Cryptography and Security,
  2006.

\bibitem{Aydre2021DifferentiallyPQ}
S.~Ayd{\"o}re, W.~Brown, M.~Kearns, K.~Kenthapadi, L.~Melis, A.~Roth, A.~Siva,
  Differentially private query release through adaptive projection, in:
  International Conference on Machine Learning, 2021.

\bibitem{Wang2020ContinuousRO}
T.~Wang, J.~Q. Chen, Z.~Zhang, D.~Su, Y.~Cheng, Z.~Li, N.~Li, S.~Jha,
  Continuous release of data streams under both centralized and local
  differential privacy, Proceedings of the 2021 ACM SIGSAC Conference on
  Computer and Communications Security (2020).

\bibitem{Maruseac2020PrecisionEnhancedDM}
M.~Maruseac, G.~Ghinita, Precision-enhanced differentially-private mining of
  high-confidence association rules, IEEE Transactions on Dependable and Secure
  Computing 17 (2020) 1297--1309.

\bibitem{Wang2018LocallyDP}
T.~Wang, N.~Li, S.~Jha, Locally differentially private frequent itemset mining,
  2018 IEEE Symposium on Security and Privacy (SP) (2018) 127--143.

\bibitem{Triastcyn2019BayesianDP}
A.~Triastcyn, B.~Faltings, Bayesian differential privacy for machine learning,
  in: International Conference on Machine Learning, 2019.

\bibitem{Zheng2020ProtectingDB}
H.~Zheng, Q.~Ye, H.~Hu, C.~Fang, J.~Shi, Protecting decision boundary of
  machine learning model with differentially private perturbation, IEEE
  Transactions on Dependable and Secure Computing 19 (2020) 2007--2022.

\bibitem{Jiang2023ApplicationsOD}
H.~Jiang, J.~Pei, D.~Yu, J.~Yu, B.~Gong, X.~Cheng, Applications of differential
  privacy in social network analysis: A survey, IEEE Transactions on Knowledge
  and Data Engineering 35 (2023) 108--127.

\bibitem{Dwork2006CalibratingNT}
C.~Dwork, F.~McSherry, K.~Nissim, A.~D. Smith, Calibrating noise to sensitivity
  in private data analysis, in: Theory of Cryptography Conference, 2006.

\bibitem{McSherry2009PrivacyIQ}
F.~McSherry, Privacy integrated queries: an extensible platform for
  privacy-preserving data analysis, Proceedings of the 2009 ACM SIGMOD
  International Conference on Management of data (2009).

\bibitem{Proserpio2012CalibratingDT}
D.~Proserpio, S.~Goldberg, F.~McSherry, Calibrating data to sensitivity in
  private data analysis, Proc. VLDB Endow. 7 (2012) 637--648.

\bibitem{Nissim2007SmoothSA}
K.~Nissim, S.~Raskhodnikova, A.~D. Smith, Smooth sensitivity and sampling in
  private data analysis, in: Symposium on the Theory of Computing, 2007.

\bibitem{Chaudhuri2017ApproximateQP}
S.~Chaudhuri, B.~Ding, S.~Kandula, Approximate query processing: No silver
  bullet, Proceedings of the 2017 ACM International Conference on Management of
  Data (2017).

\bibitem{Haas1999RippleJF}
P.~J. Haas, J.~M. Hellerstein, Ripple joins for online aggregation, in: ACM
  SIGMOD Conference, 1999.

\bibitem{Li2016WanderJO}
F.~Li, B.~Wu, K.~Yi, Z.~Zhao, Wander join: Online aggregation via random walks,
  Proceedings of the 2016 International Conference on Management of Data
  (2016).

\bibitem{Zhao2018RandomSO}
Z.~Zhao, R.~Christensen, F.~Li, X.~Hu, K.~Yi, Random sampling over joins
  revisited, Proceedings of the 2018 International Conference on Management of
  Data (2018).

\bibitem{DBLP:conf/sigmod/DobraGGR02}
A.~Dobra, M.~N. Garofalakis, J.~Gehrke, R.~Rastogi,
  \href{https://doi.org/10.1145/564691.564699}{Processing complex aggregate
  queries over data streams}, in: M.~J. Franklin, B.~Moon, A.~Ailamaki (Eds.),
  Proceedings of the 2002 {ACM} {SIGMOD} International Conference on Management
  of Data, Madison, Wisconsin, USA, June 3-6, 2002, {ACM}, 2002, pp. 61--72.
\newblock \href {https://doi.org/10.1145/564691.564699}
  {\path{doi:10.1145/564691.564699}}.
\newline\urlprefix\url{https://doi.org/10.1145/564691.564699}

\bibitem{charikar2002finding}
M.~Charikar, K.~Chen, M.~Farach-Colton, Finding frequent items in data streams,
  in: International Colloquium on Automata, Languages, and Programming,
  Springer, 2002, pp. 693--703.

\bibitem{cormode2005improved}
G.~Cormode, S.~Muthukrishnan, An improved data stream summary: the count-min
  sketch and its applications, Journal of Algorithms 55~(1) (2005) 58--75.

\bibitem{DBLP:journals/pvldb/VengerovMZC15}
D.~Vengerov, A.~C. Menck, M.~Za{\"{\i}}t, S.~Chakkappen,
  \href{http://www.vldb.org/pvldb/vol8/p1530-vengerov.pdf}{Join size estimation
  subject to filter conditions}, Proc. {VLDB} Endow. 8~(12) (2015) 1530--1541.
\newblock \href {https://doi.org/10.14778/2824032.2824051}
  {\path{doi:10.14778/2824032.2824051}}.
\newline\urlprefix\url{http://www.vldb.org/pvldb/vol8/p1530-vengerov.pdf}

\bibitem{DBLP:conf/sigmod/0001WYZ16}
F.~Li, B.~Wu, K.~Yi, Z.~Zhao,
  \href{https://doi.org/10.1145/2882903.2915235}{Wander join: Online
  aggregation via random walks}, in: F.~{\"{O}}zcan, G.~Koutrika, S.~Madden
  (Eds.), Proceedings of the 2016 International Conference on Management of
  Data, {SIGMOD} Conference 2016, San Francisco, CA, USA, June 26 - July 01,
  2016, {ACM}, 2016, pp. 615--629.
\newblock \href {https://doi.org/10.1145/2882903.2915235}
  {\path{doi:10.1145/2882903.2915235}}.
\newline\urlprefix\url{https://doi.org/10.1145/2882903.2915235}

\bibitem{DBLP:journals/pvldb/KimBPIMR15}
A.~Kim, E.~Blais, A.~G. Parameswaran, P.~Indyk, S.~Madden, R.~Rubinfeld,
  \href{http://www.vldb.org/pvldb/vol8/p521-kim.pdf}{Rapid sampling for
  visualizations with ordering guarantees}, Proc. {VLDB} Endow. 8~(5) (2015)
  521--532.
\newblock \href {https://doi.org/10.14778/2735479.2735485}
  {\path{doi:10.14778/2735479.2735485}}.
\newline\urlprefix\url{http://www.vldb.org/pvldb/vol8/p521-kim.pdf}

\bibitem{0Probability}
W.~Hoeffding, Probability Inequalities for Sums of Bounded Random Variables
  Author ( s ): Source : Journal of the American Statistical Association , Vol
  . 58 , No . 301 ( Mar ., 1963 ), pp . 13- Published by : American Statistical
  Association Stable U, Encyclopedia of Statistical Sciences.

\end{thebibliography}
\end{document}